\documentstyle[11pt,amsmath,amssymb,epsf,epic,cite]{article}
\numberwithin{equation}{section}

\newcommand{\ra}{\rightarrow}

\newcommand{\bra}{\langle}
\newcommand{\ket}{\rangle}
\newcommand{\be}{\begin{equation}}
\newcommand{\ee}{\end{equation}}
\newcommand{\bea}{\begin{eqnarray}}
\newcommand{\eea}{\end{eqnarray}}

\newcommand{\e}{{\rm e}}

\renewcommand{\d}{{\rm d}}

\newcommand{\av}[1]{\left\langle{#1}\right\rangle}

\newcommand{\grintl}{[\kern-.18em [}
\newcommand{\grintr}{]\kern-.18em ]}

\newcounter{resultcounter}[section]

\newtheorem{thm}[resultcounter]{Theorem}
\newtheorem{lem}[resultcounter]{Lemma}
\newtheorem{prop}[resultcounter]{Proposition}

\newtheorem{definition}[resultcounter]{Definition}

\def\bed{\begin{definition}}
\def\eed{\end{definition}}

\newcommand{\C}{{\mathbb C}}

\newcommand{\E}{{\mathbb E}}

\newcommand{\r}{{\rm R}}
\newcommand{\s}{{\rm S}}

\renewcommand{\i}{{\rm i}}

\def\qed{\hfill $\Box$\medskip}

\newcommand{\fer}[1]{(\ref{#1})}
\newcommand{\scalprod}[2]{\left\langle {#1}, {#2}\right\rangle}
\newcommand{\bbbone}{\mathchoice {\rm 1\mskip-4mu l} {\rm 1\mskip-4mu l}
{\rm 1\mskip-4.5mu l} {\rm 1\mskip-5mu l}}

\renewcommand{\r}{{\rm R}}

\begin{document}


\title{Representations of canonical commutation relations describing infinite coherent states}

\author{{\small By}\\
	\ \\
	Alain Joye\footnote{ 
		Institut Fourier, UMR 5582 du CNRS, 
		Universit\'e Grenoble I, BP 74, 38402 Saint-Martin d'H\`eres, France; alain.joye@ujf-grenoble.fr;  https://www-fourier.ujf-grenoble.fr/$\sim$joye/ } \ \  and \ \ Marco Merkli\footnote{Department of
		Mathematics and Statistics, Memorial University of Newfoundland, St. John's, NL, Canada, A1C 5S7; merkli@mun.ca; http://www.math.mun.ca/$\sim$merkli/}}

\maketitle

\begin{abstract}
We investigate the infinite volume limit of quantized photon fields in multimode coherent states. We show that for states containing a continuum of coherent modes, it is mathematically and physically natural to consider their phases to be random and identically distributed. The infinite volume states give rise to Hilbert space representations of the canonical commutation relations which we construct concretely. In the case of random phases, the representations are random as well and can be expressed with the help of It\^o stochastic integrals. 

We analyze the dynamics of the infinite state alone and the open system dynamics of small systems coupled to it. We show that under the free field dynamics, initial phase distributions are driven to the uniform distribution. We demonstrate that coherences in small quantum systems, interacting with the infinite coherent state, exhibit Gaussian time decay. The decoherence is qualitatively faster than the one caused by infinite thermal states, which is known to be exponentially rapid only. This emphasizes the classical character of coherent states.
\end{abstract}

\section{Introduction}

Coherent states play an important role in the theory and experiment of quantum mechanics. They have first been discovered by Schr\"odinger in 1926  as quantum states which behave in many respects like classical  ones \cite{Schroedinger}. Fourty years after, Glauber \cite{Glauber1,Glauber2} realized that these states are particularly suited to describe optical coherence, hence their name. Their use in modern quantum optics is now ubiquitous  \cite{MandelWolf, GardinerZoller, Schlosshauer} and more recently, coherent states have been used successfully in quantum information experiments (for instance in implementations of quantum key distribution \cite{Grosshans}).

In this paper, we analyze the infinite volume limit, or  thermodynamic limit, of the quantized radiation field in a multimode coherent state. We start with the field confined to a box $\Lambda\subset{\mathbb R}^3$. The field modes (or, momenta of the associated particles) are quantized and take only discrete values. To each discrete mode $k$ is associated a creation and an annihilation operator $a^*_k$ and $a_k$, respectively, satisfying the usual canonical commutation relations $[a_k,a^*_\ell]=\delta_{k,\ell}$ (Kronecker symbol). To each mode $k$ corresponds as well a family of coherent states $|\alpha\ket_k$, indexed by $\alpha\in\mathbb C$. They are defined to be normalized eigenvectors of the annihilation operator $a_k$, satisfying $a_k|\alpha\ket_k = \alpha |\alpha\ket_k$. The state $|\alpha\ket_k$ does not have a definite number of particles as it is not an eigenvector of the number operator $\widehat N=\sum_k a^*_ka_k$. The average of $\widehat N$ in $|\alpha\ket_k$ is $|\alpha|^2$. We set $\alpha=|\alpha|\e^{\i\theta}$ and call $\theta$ the phase of the coherent state. The state $|\alpha\ket_k$ can be obtained by applying the {\em displacement operator} $D_k(\alpha)=\e^{\alpha a^*_k -\bar \alpha a_k}$ to the vacuum vector $\Omega$ of the quantum field, $|\alpha\ket_k=D_k(\alpha)\Omega$. This is a single mode coherent state. If one selects $N$ (discrete) modes $k_1,\ldots,k_N$, the associated {\em multimode coherent states} are $D_{k_1}(\alpha_1)\cdots D_{k_N}(\alpha_N)\Omega$, for any choice of $\alpha_j\in \mathbb C$, $j=1,\ldots,N$.

It is well known that the electromagnetic radiation generated by a classical current is a multimode coherent state of this form, and so is the light produced by a laser in certain regimes \cite{MandelWolf, MartinRothen}. One of the physical motivations to consider the infinite volume limit is that the radiation quantum field provides an environment (``reservoir") which is extremely large, spatially, relative to objects of interest that are placed in this environment. ``Object-environment systems" are called open quantum systems. The archetypical example is a spin $1/2$ (a ``qubit") interacting with an environment (photons, phonons). The infinite nature of the reservoir causes irreversible dynamical effects in the spin, such as thermalization, decoherence and disentanglemet (in presence of several spins). On the mathematical side, taking the infinite volume limit is interesting in its own right, as it uncovers new Hilbert space representations of the canonical commutation relations. Indeed, in 1963, Araki and Woods \cite{AW} considered the infinite volume limit of ``Fock states" having the form $a^*(f_1)\cdots a^*(f_N)\Omega$, which describe thermal equilibrium as well as condensate states of the field. They found the famous Araki-Woods representations of the canonical commutation relations. These representations, as well as their  fermionic counterparts, the Araki-Wyss representations \cite{AWy},  have proven to be an important tool in the mathematical analysis of open quantum systems close to thermal equilibrium.

In an effort to be able to handle  reservoirs, that is, spatially infinitely extended systems with non-vanishing particle density, which are physically different from thermal ones, we tackle here the question of coherent reservoirs. As mentioned above, they describe the electromagnetic field  created by classical sources or by lasers. 

\medskip

{\bf Outline of main results.} We start off with the quantized radiation field in a finite box $\Lambda\subset{\mathbb R}^3$,  pick modes $k_1,\ldots,k_N$ and consider the multimode coherent state $D_{k_1}(\alpha_1)\cdots D_{k_N}(\alpha_N)\Omega$. What happens to this state when the size of the box increases, $\Lambda\rightarrow{\mathbb R}^3$ ? 
As $\Lambda$ changes, so do the eigenmodes of the quantum field, and in the limit of infinite volume, the values of the modes becomes a continuum, $k\in{\mathbb R}^3$. Any state $\omega$ of the radiation field (in finite or infinite volume) is determined uniquely by its {\em expectation functional} $E(f)=\omega(W(f))$, where $W(f)$ is the {\em Weyl operator} smoothed out with a test function $f$. Convergence of a sequence of states $\omega_{\Lambda_n}$ (with $\Lambda_n\rightarrow {\mathbb R}^3$) is then meant as convergence of the associated $E_n(f)$, for all $f$. 
\medskip

{\bf (A) Thermodynamic limit.} We look at two basic scenarios.
\begin{itemize}
	\item {\em Infinite volume limit for $N$ fixed modes}: Fix $k_1,\ldots,k_N\in{\mathbb R}^3$ and consider the limit $\Lambda\rightarrow{\mathbb R}^3$ while keeping the particle densities $\rho_j=|\alpha_j|^2/|\Lambda|$, and the phases $\theta_j\in S^1$, $j=1,\ldots,N$, fixed.
\end{itemize}
	We show that the $N$-mode state (i.e., the associated expectation functional) has a thermodynamic limit and calculate the latter explicitly in Proposition \ref{proposition1}. The result depends, of course, on the densities $\rho_j$ and the phases $\theta_j$, $j=1,\ldots N$.
	
\begin{itemize}
	\item {\em Infinite volume limit and continuous mode limit}: Fix a mode density distribution $\rho(k)$, meaning that $\rho(k)\d^3k$ is the spatial density of particles (number of particles per unit volume in direct space) having momenta in an infinitesimal volume $\d^3k$ around $k$. Consider the infinite volume limit of the state associated to $\rho(k)$.
\end{itemize}
The ``naive" infinite volume and continuous mode limit of the multimode coherent state (i.e., its expectation functional) does not exist, see \eqref{div}. However, the state obtained by mixing the phases of the $N$ modes in an identical and independent way according to a measure $\mu$ on the unit circle, does converge. The limit depends on the mode density distribution $\rho(k)$ as well as on the phase distribution $\mu$, see Proposition \ref{proposition2}. It is interesting to note that in laser  experiments, the intensity of the laser field (measured by $\rho(k)$) is observed to be almost free of fluctuations due to saturation properties of the laser. However, the phases are subject to fluctuations and drift randomly. The theoretical framework describing these properties is called {\em the randomly phased laser model} (see e.g. \cite{MandelWolf}, paragraph 11.8.6). The fluctuation of the phases is precisely what makes the infinite volume and continuous mode limit exist in our analysis. Considering the phase mixed state is the same as taking an expectation of random phases, and the above mentioned result (Proposition \ref{proposition2}) is equivalent to the convergence of the expectation of the state with random phases. We then ask if a stronger convergence holds. Consider the phases to be iid, distributed according to a measure $\mu$ on $S^1$. We show in Proposition \ref{cltprop} that the corresponding random phase state {\em converges in distribution}. The convergence is in essence due to the central limit theorem. The random part of the limit expectation functional $E(f)$ is given by $\e^{\i {\rm Re}\chi(f)}$, where $\chi(f)$ is a complex valued random variable which we construct as a suitable It\^o stochastic integral. It has the property ${\rm Re}\chi(f)\sim {\cal N}(0,\sigma_\mu(f)^2)$, where the latter is a (real) normal random variable with mean zero and variance $\sigma_\mu(f)^2$ which depends on the phase distribution $\mu$. This result is the content of Theorem \ref{thm1}. We point out that only the second Fourier moment $\widehat\mu(2)$ (c.f. \eqref{mumoments}) of the phase distribution $\mu$ enters the infinite volume expression.

\medskip
{\bf (B) Hilbert space representation.} In Section \ref{hilbertsection} we construct explicit Gelfand-Naimark-Segal (GNS) Hilbert spaces of the three infinite volume states constructed before, namely, the finite modes state, the phase averaged state and the random phases state. Those representations are given in Theorem \ref{repthm}. They all are regular representations of the canonical commutation relations, hence defining (represented) field-, creation- and annihilation operators (Proposition \ref{fieldopsprop}).

\medskip
{\bf (C) Dynamics of the infinite coherent state.} The Heisenberg dynamics of the field is a Bogoliubov transformation $W(f)\mapsto W(\e^{\i t \varepsilon }f)$, where $t$ is time and (in Fourier space) $\varepsilon=\varepsilon(k)$ is the dispersion relation. For the infinite volume state with $N$ fixed coherent modes $k_j$ and phases $\theta_j$, the effect of the dynamics is to rotate the phases, $\theta_j\mapsto \theta_j-t\varepsilon(k_j)$, making the infinite $N$ mode coherent state a quasiperiodic function of time. However, the situation of the mixed phase state and of the random phases state is quite different. We show in Proposition \ref{propconv} that the dynamics drives those states into a final state, which corresponds to the {\em uniform distribution} of the phases. More precisely, starting off with the infinite volume  state with phase distribution $\mu$ on the circle (and continuous mode distribution $\rho$), as time moves on, the system converges to the infinite volume state with phase distribution $\d\mu(\theta) = \d\theta/2\pi$ (and the same continuous mode distribution $\rho$). In this sense, the uniform phase distribution is the  stable one.

\medskip
{\bf(D) Coupling to an open quantum system. } In Section \ref{opensystdyn} we consider an $N$-level quantum system in contact with the infinite volume coherent reservoir having uniformly randomly distributed phases (the dynamically stable reservoir state as explained in point (C) above). The coupling between the $N$-level system and the reservoir is {\em energy conserving}, meaning that the interaction term in the Hamiltonian describing the coupled evolution commutes with the system Hamiltonian. Such couplings are often considered to describe so-called {\em phase decoherence} without energy exchange \cite{PSE,JZKGKS,MSB}. This model is {\em explicitly solvable} in the sense that we can calculate the exact density matrix of the $N$-level system at all times (c.f. \eqref{deco1}, \eqref{deco3}). We show that the expectation of the off-diagonal density matrix elements (in the energy basis) have {\em Gaussian} time decay due to the coupling with the coherent reservoir. This is a striking difference relative to the thermal reservoir case, where this decay is only exponential. We conclude that a small system placed in a coherent reservoir undergoes {\em much faster decoherence} than in a  thermal environment.\footnote{The name ``coherent states" refers to the quantum field and is motivated by the fact that correlation functions of the field factorize in those states, which is the same as for classical coherent fields \cite{MandelWolf}. On the other hand ``decoherence" of a quantum system is an entirely different notion, which refers to a system losing quantum correlations and becoming close to a classical one \cite{JZKGKS}.} This very rapid loss of ``quantumness" (encoded by coherence of the small system) is yet another manifestation that coherent states behave to some extent classically.

\medskip

\section{Setup and main results}

We consider non-interacting quantum particles confined to a box of sidelength $L$ in $d$ dimensions, 
$$
\Lambda=[-L/2,L/2]^d\subset{\mathbb R}^d.
$$
The wave function of a single particle is an element of $L^2(\Lambda,\d x)$ with periodic boundary conditions. The space of pure states of the system of particles is the symmetric Fock space \cite{BRI}
$$
{\cal F}\equiv {\cal F}\big(L^2(\Lambda,\d x)\big) = \bigoplus_{n\geq 0} L^2_{\rm symm}(\Lambda^n,\d^nx).
$$
A state is then given by $\psi=\oplus_{n\geq 0}\psi^{(n)}$, where $\psi^{(n)}(x_1,\ldots,x_n)$ is a symmetric function of $n$ variables $x_j\in{\mathbb R}^d$ with periodic boundary conditions in $\Lambda^n$. The summand $n=0$ of the Fock space is called the vacuum sector, it equals ${\mathbb C}$ and is spanned by the vacuum vector $\Omega$ (such that $\Omega^{(0)}=1$, $\Omega^{(n)}=0$ for $n\geq 1$).

The dynamics obeys the Schr\"odinger equation
\begin{equation*}
\psi_t = \e^{-\i Ht}\psi_0,
\end{equation*}
where the self-adjoint Hamiltonian is the second quantization of a one-body Hamilton operator. For photons (massless relativistic particles), $H$ is the square root of the Laplace operator,
$$
(H\psi)^{(n)}(x_1,\ldots,x_n)= \sum_{j=1}^n \sqrt{-\Delta_{x_j}} \psi^{(n)}(x_1,\ldots,x_n),
$$
understood as a self-adjoint operator with periodic boundary conditions. 

The {\em creation operator} $a^*(f)$ is defined for $f\in L^2(\Lambda,\d x)$ by
$$
(a^*(f)\psi)^{(n)}(x_1,\ldots,x_n) = \sqrt{n} \, {\cal S}f(x_1)\psi^{(n-1)}(x_2,\ldots,x_n),
$$
where $\cal S$ is the operator of symmetrization over the variables $x_1,\ldots,x_n$. The adjoint of the creation operator is the {\em annihilation operator} $a(f)$, given by
$$
 (a(f)\psi)^{(n)}(x_1,\ldots,x_n) = \sqrt{n+1}\, \int_{\Lambda} \bar f(x)\psi^{(n+1)}(x,x_1,\ldots,x_n) \d x.
$$
The self-adjoint {\em field operator} is defined by
$$
\Phi(f) =\frac{1}{\sqrt{2}} \big( a^*(f)+a(f)\big)
$$
and the unitary {\em Weyl operators} is 
$$
W(f) =\e^{\i\Phi(f)}.
$$
It is practical to introduce the operator valued distributions $a^*(x)$, $a(x)$ by setting
$$
a^*(f) = \int_\Lambda f(x)a^*(x)\d x,\qquad a(f) = \int_\Lambda \bar f(x)a(x)\d x.
$$
The {\em canonical commutation relations} take the following equivalent forms:
\begin{eqnarray}
{}[a(f),a^*(g)] &=& \scalprod{f}{g}\nonumber \\
{}[a(x),a^*(y)] &=& \delta(x-y)\nonumber\\
{}[\Phi(f),\Phi(g)] &=& \i\, {\rm Im}\scalprod{f}{g}\nonumber\\
W(f)W(g)&=& \e^{-\frac{\i}{2}{\rm Im}\scalprod{f}{g}} W(f+g).
\label{weylccr}
\end{eqnarray}
Let $\psi$ be a normalized vector in $\cal F$. It determines an expectation functional, defined by $E(f):=\scalprod{\psi}{W(f)\psi}$, $f\in L^2(\Lambda,\d x)$. Conversely, any functional $E: L^2(\Lambda,\d x)\rightarrow\mathbb C$ satisfying the three conditions
\begin{itemize}
\item[(E1)] $E(0)=1$ 
\item[(E2)] $\overline{E(f)}=E(-f)$
\item[(E3)] $\sum_{k,k'=1}^K z_k\overline{z_{k'}} \e^{\frac{\i}{2}{\rm Im}\,\scalprod{f_{k}}{f_{k'}}} E(f_k-f_{k'}) \geq 0$, for all $K\geq 1$, $z_k\in\mathbb C$, $f_k\in L^2(\Lambda,\d x)$
\end{itemize}
determines a state $\omega$ on the $C^*$-algebra generated by the Weyl operators by the relation $\omega(W(f))=E(f)$, see for instance  \cite{MIdeal}.

\smallskip

To define {\em coherent} states of the particles in the volume $\Lambda$ it is convenient to pass to the momentum space representation, the Fourier transformation of the Fock space $\cal F$. There coherent states take a simple form, while their expression in position space (as vectors in $\cal F$) is more cumbersome.

\subsection{Momentum space representation}

The single-particle Hilbert space $L^2(\Lambda,\d x)$ is unitarily equivalent to $l^2(\frac{2\pi}{L}{\mathbb Z}^d)$ via the Fourier transform ${\frak F}:  L^2 (\Lambda,\d x)\rightarrow  l^2(\frac{2\pi}{L}{\mathbb Z}^d)$,
\begin{equation}
({\frak F} f)(k) = \widehat{f}_k= L^{-d/2} \int_\Lambda \e^{-\i kx} f(x) \d x,
\label{1}
\end{equation}
having the inverse
$$
({\frak F}^{-1}\widehat{f})(x) = L^{-d/2} \sum_{k\in\frac{2\pi}{L}{\mathbb Z}^d} \e^{\i kx} \widehat{f}_k.
$$
Here, $kx$ is the dot product $k\cdot x$ and the factors $L^{-d/2}$ guarantee that ${\frak F}$ is unitary. Accordingly, the Fock space $\cal F$ is unitarily equivalent to its momentum version
$$
\widehat{\cal F}\equiv {\cal F}\big(l^2(\textstyle\frac{2\pi}{L}{\mathbb Z}^d)\big) = \bigoplus_{n\geq 0} \Big(l^2(\frac{2\pi}{L}{\mathbb Z}^d)\Big)^{{\otimes}^n_{\rm symm}}.
$$
Let $\Omega$ and $\widehat\Omega$, be the vacua of the Fock spaces $\cal F$ and $\widehat{\cal F}$, respectively. The unitary map between the Fock spaces is given by the natural lifting of $\frak F$,
$$
{\frak F} a^*(f_1)\cdots a^*(f_\ell)\Omega = a^*(\widehat f_1)\cdots a^*(\widehat f_\ell)\widehat\Omega.
$$
Accordingly, the creation operators transform as ${\frak F}a^*(f){\frak F}^{-1} = a^*(\widehat{f} )$. We write
$$
a^*(\widehat{f} )=\sum_{k\in \frac{2\pi}{L}{\mathbb Z}^d} \widehat{f}_k a^*_k
$$
and so
$$
a^*(\widehat f) = \int_\Lambda f(x) \Big(L^{-d/2} \sum_{k\in \frac{2\pi}{L}{\mathbb Z}^d} \e^{-\i kx} a^*_k\Big)\d x.
$$
Comparing this with ${\frak F}a^*(f){\frak F}^{-1}=\int_\Lambda f(x) {\frak F}a^*(x){\frak F}^{-1}\d x$ yields the relations
$$
{\frak F}a^*(x){\frak F}^{-1} = L^{-d/2}\sum_{k\in \frac{2\pi}{L}{\mathbb Z}^d} \e^{-\i kx}  a^*_k, \qquad  {\frak F}^{-1}a^*_k\, {\frak F} = L^{-d/2} \int_\Lambda \e^{\i kx} a^*(x)\d x.
$$
The field- and Weyl operators are transported to the momentum space as
$$
{\frak F} \Phi(f){\frak F}^{-1} = \Phi(\widehat f)\qquad\mbox{and}\qquad {\frak F} W(f){\frak F}^{-1} = W(\widehat f),
$$
where $\Phi(\widehat f) = \frac{1}{\sqrt{2}}\sum_{k\in\frac{2\pi}{L}{\mathbb Z}^d}(\widehat f_k a^*_k +\overline{\widehat f_k} a_k)$ and 
 $W(\widehat f)=\e^{\i \Phi(\widehat f)}$.

\subsection{$N$-mode coherent states in finite volume} 

The {\em coherent state} associated to the collection of $N$ modes $k'_1,\ldots,k_N'\in \frac{2\pi}{L}{\mathbb Z}^d$ and $N$ complex numbers $\alpha_1,\ldots,\alpha_N$ is the normalized vector 
\begin{equation}
\widehat\Psi = \e^{\sum_{j=1}^N \alpha_j a^*_{k'_j} -\bar\alpha_j a_{k'_j}} \widehat \Omega  \in \widehat{\cal F}.
\label{cohstate}
\end{equation}
The {\em expectation functional} of the coherent state \fer{cohstate} is defined by
\begin{equation}
E^\Lambda_N(f) =\scalprod{\widehat\Psi}{W(\widehat f)\widehat\Psi},
\label{mf1}
\end{equation}
for all $\widehat f\in l^2(\frac{2\pi}{L}{\mathbb Z}^d)$. The operator $a^*_{k'_j}a_{k'_j}$ is the number operator of the mode $k'_j\in\frac{2\pi}{L}{\mathbb Z}^d$. Its average in the coherent state is 
$$
\scalprod{\widehat \Psi}{a^*_{k'_j}a_{k'_j}\widehat\Psi} =|\alpha_j|^2
$$
and can be interpreted as the intensity of the mode in question.

\subsection{Infinite volume and continuous mode limits, random phases}

\subsubsection{Infinite volume}

The momenta in the finite-volume coherent state are given by $k'_j=2\pi n_j/L$, for $n_j\in{\mathbb Z}^d$. As $L$ increases, the spacing of the momenta becomes increasingly small. Let now $k_1,\ldots,k_N\in {\mathbb R}^d$ be $N$ arbitrary (`continuous') momenta and let $n_j=n_j(L)\in{\mathbb Z}^d$ be such that $k'_j(L)= 2\pi n_j(L)/L$ satisfies $\lim_{L\rightarrow\infty}k'_j(L)=k_j$, $j=1,\ldots,N$. We want to take the thermodynamic limit of \eqref{mf1},
\begin{equation}
\lim_{L\rightarrow\infty} E^\Lambda_N(f) \equiv E_N(f).
\label{mf2}
\end{equation}
This means we take $k_j'=k'_j(L)$ and $L\rightarrow\infty$, while keeping fixed the particle densities $\rho_j\geq 0$ which count the (average) number of particles in mode $k'_j$ per unit volume, for $j=1,\ldots,N$. In other words, we have $|\alpha_j|^2 = L^d\rho_j$, or
\begin{equation}
\label{alpha}
\alpha_j(L) =L^{d/2}\sqrt{\rho_j}\  \e^{\i \theta_j},
\end{equation}
where $\theta_j$ is the phase of the complex number $\alpha_j$.

\begin{prop}[Thermodynamic limit for $N$ modes]
\label{proposition1}
Let $k_1,\ldots,k_N\in\mathbb R$ and $\rho_1,\ldots,\rho_N\geq 0$ be arbitrary momenta and arbitrary particle densities and suppose that $f\in L^1({\mathbb R}^d,\d x)\cap L^2({\mathbb R}^d,\d x) $. Then the limit \fer{mf2} exists and 
\begin{equation}
E_N(f) = E_{{\rm Fock}}(f)\  \e^{ \i\, {\rm Re}\sum_{j=1}^N  \e^{-\i\theta_{j}} \sqrt{2\rho_j}\ \widehat f(k_j)},
\label{6}
\end{equation}
where $E_{{\rm Fock}}(f) = \e^{-\frac14 \|f\|^2}$ and $\widehat f(k)=\int_{{\mathbb R}^d}\e^{-\i k x} f(x)\d x$.
\end{prop}

Here, $E_{{\rm Fock}}(f)$ is the {\em Fock expectation functional}, determined by the vacuum state, 
$$
E_{\rm Fock}(f) = \scalprod{\Omega}{W(f)\Omega} = \e^{-\frac14 \|f\|^2_2} =  \e^{-\frac14 (2\pi)^{-d}\|\widehat f \|^2_2}.
$$
Here, $\| \cdot \|_2$ is the $L^2$-norm (of functions of $k\in{\mathbb R}^d$ or $x\in{\mathbb R}^d$), 
$$
\|f\|^2_2 = \int_{{\mathbb R}^d} |f(x)|^2 \d x \mbox{\quad and\quad} \|\widehat f \|^2_2 = \int_{{\mathbb R}^d} |\widehat f (k)|^2 \d k.
$$

{\em Remark.\ } By adopting the definition  $\widehat f(k)=\int_{{\mathbb R}^d}\e^{-\i k x} f(x)\d x$ of the Fourier transform (see Proposition \ref{proposition1}), we obtain  $f(x)=(2\pi)^{-d} \int_{{\mathbb R}^d}\e^{\i k x} \widehat f (k)\d k$ and 
$$
\|f\|^2_2 = (2\pi)^{-d}\|\widehat f\|^2_2.
$$

\subsubsection{Continuous modes}

We are now interested in an infinite-volume coherent state which contains a contiuum of modes. One may perform the infinite-volume limit and the continuous mode limit simultaneously, or one can take the continuous mode limit of \fer{6}. The result is the same and we do the latter (see Section \ref{sameresultsection}). Let $\rho(k)$ be a prescribed mode density distribution. That is, given a cube $I\subset {\mathbb R}^d$, the integral $\int_I \rho(k)\d k$ is the spatial density of particles in the infinite volume state, having momenta in $I$. Consider $\rho$ to be supported in a finite cube $[-R,R]^d$.  Discretize the cube by taking an $N$ (large) and setting
\begin{equation}
\label{kj}
k_j=(-R+j_1\frac{2R}{N},\ldots, -R+j_d\frac{2R}{N})\in {\mathbb R}^d,
\end{equation}
where $j_1,\ldots,j_d\in\{1,2,\ldots, N\}$. Here, we view $j=(j_1,\ldots,j_d)$ as a multi-index.

In the previous considerations, $\rho_j$ was the number of particles per unit volume having momentum $k_j$. Since $\rho(k)$ is, by definition, the number of particles per unit volume in space and per momentum volume $\d k$, we have $\rho(k_j)=\frac{\rho_j}{\Delta k_j}$, with $\Delta k_j=(2R/N)^d$. Consequently, the sum in the phase of the infinite volume expectation functional \fer{6} is
\begin{equation}
\sum_{j\in \{1,\ldots,N\}^d}  \e^{-\i\theta_j} \sqrt{2\rho_j}\ \widehat f(k_j) = (2R/N)^{d/2} \sum_{j\in \{1,\ldots,N\}^d} \e^{-\i\theta(k_j)}\sqrt{2\rho(k_j)} \,\widehat f(k_j).
\label{mf3}
\end{equation}
Here, $\theta(k)$ is a function we can choose, which determines the phase of the mode $k$. For $N$ large, \fer{mf3} equals approximately  
\begin{equation}
\label{div}
(N/2R)^{d/2} \int_{[-R,R]^d} \e^{-\i\theta(k)}\sqrt{2\rho(k)}\, \widehat f(k) \d k \ \ \sim \ \ N^{d/2},
\end{equation}
which diverges as $N\rightarrow\infty$. It follows that the infinite volume discrete mode expectation functional $E_N(f)$, \fer{6}, does not have a continuous mode limit in this simple sense.

\subsubsection{Phase mixture} 

Suppose the phases in the finite-volume coherent state \eqref{cohstate} are not fixed, but distributed independently according to probability measures $\d\mu_j$ on the circle. For given phases, denote the coherent state by $\widehat \Psi_{\theta_1,\ldots,\theta_N}$. The mixed state is given by the density matrix
$$
\widehat\varrho = \int_0^{2\pi}\d\mu_1(\theta_1)\cdots\int_0^{2\pi}\d\mu_{N}(\theta_{N}) \ |\widehat\Psi_{\theta_1,\ldots,\theta_{N}} \rangle\langle\widehat \Psi_{\theta_1,\ldots,\theta_{N}}|.
$$
It follows immediately from Proposition \ref{proposition1} that the associated  expectation functional $\av{E_N^\Lambda}(f) = {\rm Tr}\big(\widehat\varrho\, W(\widehat f)\big)$ has the infinite volume limit
\begin{equation}
\av{E_N}(f) = E_{\rm Fock}(f) \prod_{j=1}^{N} \int_0^{2\pi} \d\mu_j(\theta)\, \e^{\i\,{\rm Re}\,\e^{-\i\theta} \sqrt{2\rho_j}\, \widehat f(k_j)}.
\label{ef1}
\end{equation}
Here, we use the notation $\av{\ }$ to indicate that we have taken the average over phases (phase mixture). 
{}For identically and uniformly distributed phases, $\d\mu_j=(2\pi)^{-1}\d\theta$, \fer{ef1} becomes
\begin{equation}
\av{E_N}(f) =E_{\rm Fock}(f)\, \prod_{j=1}^N J_0\big(\sqrt{2\rho_j}\, |\widehat f(k_j)|\big),
\label{ef2}
\end{equation}
where 
\begin{equation}
J_0\big(\sqrt{a^2+b^2}\big) =\int_0^{2\pi}\frac{\d\theta}{2\pi} \,\e^{-\i (a\cos\theta+b\sin\theta)}
\label{bessel}
\end{equation}
is the Bessel function.

The  continuous mode limit of the uniformly distributed, phase averaged expectation functional  is now well defined. To see why this is the case, consider the $N^d$ phases $\theta_j=\theta(k_j)$, where $k_j$ is given in \eqref{kj}. The averaged functional is (see also \eqref{mf3})
$$
\av{E_N}(f)=E_{\rm Fock}(f)\prod_{j\in\{1,\ldots,N\}^d}  \int_0^{2\pi} \frac{\d\theta}{2\pi}\, \e^{\i (2R/N)^{d/2} {\rm Re}\, \e^{-\i\theta}\sqrt{2\rho(k_j)}\, \widehat f(k_j)}.
$$
Since
\begin{equation}
\label{condmeas}
\int_0^{2\pi}\frac{\d\theta}{2\pi}\e^{-\i\theta} =0
\end{equation}
the first order term in the expansion of the integral in powers of $N^{-d/2}$ vanishes, and we have 
\begin{equation}
\label{vanterm}
\av{E_N}(f)=E_{\rm Fock}(f)\prod_{j\in\{1,\ldots,N\}^d}  \left[ 1 - \frac{(2R)^d\, \rho(k_j)|\widehat f(k_j)|^2}{2 N^d}+O(N^{-3d/2})\right].
\end{equation}
Consequently,
\begin{eqnarray*}
\lim_{N\rightarrow\infty} \ln \frac{\av{E_N}(f)}{E_{\rm Fock}(f)}  &=& -  \tfrac12 \lim_{N\rightarrow\infty} (2R/N)^{d}\sum_{j\in \{1,\ldots, N\}^d} \left[ \rho(k_j) |\widehat f(k_j)|^2 +O(N^{-d/2})\right]\nonumber\\
&=& -\tfrac12\int \rho(k)|\widehat f(k)|^2\d k.
\end{eqnarray*}
This derivation, which we have carried out for $\widehat f$ and $\rho$ continuous and compactly supported, can be extended to $\widehat f \in L^2({\mathbb R}^d)$ by a density argument.

In case the phases are identically and independently distributed according to a measure $\d\mu$ on the circle, the continuous mode limit exists if and only if $\widehat \mu(1)=0$, where
\begin{equation}
\label{mumoments}
\widehat \mu(n):=\int_0^{2\pi}\d\mu(\theta) \e^{-\i n\theta},\quad n\in\mathbb Z.
\end{equation}
This is so because $\widehat \mu(1)=0$, which is the analogue of \eqref{condmeas} above, is equivalent to the vanishing of the $N^{-d/2}$ term in \eqref{vanterm}. A little generalization of the above calculation, where $\d\mu$ was uniform,  shows the following result.

\begin{prop}
\label{proposition2}
Let $\rho(k)$ be a continuous distribution of momenta per unit spatial volume, having compact support. Suppose the phases are identically and independently distributed on $[0,2\pi]$, according to a probability measure $\mu$ satisfying $\widehat \mu(1)=0$ (see \eqref{mumoments}). Then the phase mixed expectation functional \fer{ef1} has the continuous mode limit
\begin{equation}
\label{001}
\lim_{N\rightarrow\infty} \av{E_N}(f) \equiv \av{E}(f) = E_{\rm Fock}(f) \, \e^{-\frac12\sigma_\mu(f)^2},
\end{equation}
where
\begin{equation}
\label{newvar}
\sigma_\mu(f)^2 := \int_{{\mathbb R}^d} \rho(k)\Big( |\widehat f(k)|^2 +{\rm Re} \, \{\widehat \mu(2) \widehat f(k)^2\} \Big)\d k.
\end{equation}
\end{prop}

\subsubsection{Random phases}

Considering the phases $\theta_j=\theta_j(\omega)$, $j\in\{1,\ldots,N\}^d$, to be random variables, the coherent state $\widehat\Psi=\widehat\Psi_\omega$ given in \fer{cohstate} is a random pure state defining the random expectation functional (in infinite volume, by Proposition \ref{proposition1})
\begin{equation}
E_{N,\omega}(f) = E_{\rm Fock}(f)\, e^{\i N^{-d/2} \sum_{j\in\{1,\ldots,N\}^d}\xi_j(\omega)},
\label{mf7}
\end{equation}
where
\begin{equation}
\label{xiomega}
\xi_j(\omega) = (2R)^{d/2}\sqrt{2\rho(k_j)}\,  {\rm Re}\, \e^{-\i\theta_j(\omega)} \,\widehat f(k_j).
\end{equation}
Denote by ${\mathbb E}$ the expectation over the randomness $(\omega)$. If the phases are independent and identically distributed over $[0,2\pi]$ according to $\mu$ with $\widehat \mu(1)=0$, then Proposition \ref{proposition2} states that 
$$
\lim_{N\rightarrow\infty} {\mathbb E}[E_{N,\omega}(f)] = \av{E}(f).
$$
On the other hand, we know from \eqref{div} that $E_{N,\omega}(f)$ does not converge almost everywhere. Does $E_{N,\omega}(f)$ converge in a sense that lies in between these two? The answer is given by the central limit theorem. 

\begin{prop}
\label{cltprop}
Suppose the phase distribution satsifies $\widehat \mu(1)=0$.  Then
\begin{equation}
\label{--1}
N^{-d/2} \sum_{j\in\{1,\ldots,N\}^d} \xi_j(\omega) \ \stackrel{\mathcal D}{\longrightarrow}\  {\cal N}_\omega \big(0,\sigma_\mu(f)^2\big),\quad \mbox{as $N\rightarrow\infty$},
\end{equation}
The right hand side is a normal random variable with mean zero and variance $\sigma_\mu ^2(f)$ given in \eqref{newvar}. The convergence is in distribution. 
\end{prop}

The meaning of convergence in distribution is that the distribution function 
$$
F_N(x)={\mathbb P}\left(N^{-d/2} \sum_{j\in\{1,\ldots,N\}^d} \xi_j(\omega) \leq x\right)
$$
converges pointwise to that of the normal with mean zero and variance $\sigma_\mu(f)^2$,
$$
\lim_{N\rightarrow\infty} F_N(x) = \frac{1}{\sqrt{2\pi}\, \sigma_\mu(f)}  \int_{-\infty}^x \e^{-t^2/(2\sigma_\mu(f)^2)}\d t,
$$
for all $x\in\mathbb R$.

It follows from the fact that $x\mapsto\e^{\i x}$ is bounded and continuous and \fer{mf7} that 
\begin{equation}
E_{N,\omega}(f)  \ \stackrel{\mathcal D}{\longrightarrow}\   E_\omega(f)\equiv E_{\rm Fock}(f) \,\e^{\i \,{\cal N}_\omega(0,\sigma_\mu(f)^2)},\quad \mbox{as $N\rightarrow\infty$}.
\label{mf8}
\end{equation}
Our task now is to find an explicit representation of the random variable ${\cal N}_\omega(0,\sigma_\mu(f)^2)$ such that the corresponding $E_\omega(f)$, \eqref{mf8}, defines an expectation functional. Let $X_\omega(f)$ be such a random variable. It is easily seen that $E_\omega(\cdot)$ satisfies conditions (E1)-(E3) provided that $X_\omega(-f)=-X_\omega(f)$ and $X_\omega(f_1+f_2)=X_\omega(f_1)+X_\omega(f_2)$. We are then taking $X_\omega(f)={\rm Re} \chi_\omega(f)$, where, for $f\in L^2({{\mathbb R}^d},\d x)$, we define $\chi_\omega$ as an It\^o integral
\begin{equation}
\chi_\omega(f) = \int_{{\mathbb R}^d} \d B^\omega_1(k)   S_1(k) \widehat f (k)+\i\, \int_{{\mathbb R}^d} \d B^\omega_2(k)  S_2(k) \widehat f(k),
\label{mf10}
\end{equation}
with
\begin{eqnarray}
S_1(k) &=& \sqrt{\frac{\rho(k)}{1+{\rm Re}\widehat\mu(2)}}\, (1+\widehat\mu(2))\label{s1}\\
S_2(k) &=&  \sqrt{\frac{\rho(k)}{1+{\rm Re}\widehat\mu(2)}}\, \sqrt{1-|\widehat\mu(2)|^2}.
\label{s2}
\end{eqnarray}
Here, $B_1^\omega$ and $B_2^\omega$ are two independent Brownian motions of dimension $d$, as we are going to explain below.

{\em Remarks.\ } The choices of $S_1$, $S_2$ are not unique. Since $|\widehat\mu(2)|\le 1$ one easily sees that both $S_1(k)$ and $S_2(k)$ are bounded as functions of $\widehat\mu(2)$. If ${\rm Re}\widehat\mu(2)=-1$, then one may take $S_1=\i\sqrt{\rho}$ and $S_2=\sqrt{\rho}$. This happens e.g. if $\d \mu = \tfrac12\{ \delta_{\pi/2}+\delta_{-\pi/2}\}$.

 The (almost sure, real) linearity of \eqref{mf10} in $f$ guarantees that 
\begin{equation}
E_\omega(\cdot) = E_{\rm Fock}(\cdot) \e^{\i{\rm Re}\chi_\omega(\cdot)}
\label{Eomega}
\end{equation}
is indeed an expectation functional (satisfying (E1)-(E3)). Furthermore, 
\begin{equation}
{\rm Re} \chi_\omega(f) = \int_{{\mathbb R}^d} \d B_1^\omega(k) \, {\rm Re} \{S_1(k)\widehat f(k)\} -\int_{{\mathbb R}^d} \d B_2^\omega(k) S_2(k)\, {\rm Im}\widehat f(k)
\end{equation}
is the sum of two independent normal random variables with mean zero and variances $\|{\rm Re} S_1 \widehat f\|_2^2$ and $\|S_2 {\rm Im}\widehat f\|_2^2$ (It\^o isometry, see below), whose sum equals $\sigma_\mu(f)^2$ (use \eqref{s1}, \eqref{s2}). Therefore, ${\rm Re} \chi_\omega(f) \sim {\cal N}_\omega(0, \sigma_\mu(f)^2)$, 
as desired. This shows the following result.

\begin{thm}
\label{thm1} 

Suppose  that $\widehat \mu(1)=0$. \\

{\rm \bf (1)} For all $f\in L^2({\mathbb R}^d,\d k)$,
\begin{equation}
{\rm Re}\, \chi_\omega(f) \ \sim \  {\cal N}_\omega\big(0,\sigma_\mu(f)^2\big),
\label{002}
\end{equation}
\qquad \quad \ where $\sigma_\mu(f)^2$ is given in \eqref{newvar}.

{\rm \bf (2)} Let $E_{N,\omega}(f)$ and $E_\omega(\cdot)$ be the functionals \eqref{mf7} and \eqref{Eomega}, respectively. Then 

\qquad we have,  for all $f\in L^2({{\mathbb R}^d},\d x)$,
$$
E_{N,\omega}(f)  \ \stackrel{\mathcal D}{\longrightarrow}\  E_\omega(f), \quad\mbox{as $N\rightarrow\infty$}.
$$

{\rm \bf (3)} $E_\omega(\cdot)$ satisfies (E1)-(E3) in the following sense: 

\medskip

\qquad -- $E_\omega(0)=1$ a.e.$(\omega)$

\qquad --  for all $f\in L^2({\mathbb R}^d,dx)$, $\overline{E_\omega(f)} = E_\omega(-f)$ a.e.$(\omega)$

\qquad -- for all $K\ge 1$, $z_k\in\mathbb C$, $f_k\in L^2({\mathbb R}^d,dx)$, $k=1,\ldots,K$, we have  

\qquad \quad $\sum_{k,k'=1}^K z_k\overline{z_{k'}} \e^{\frac{\i}{2}{\rm Im}\scalprod{f_k}{f_{k'}}} E_\omega(f_k-f_{k'}) \geq 0$ a.e.$(\omega)$
\end{thm}

\noindent
{\em Remark.\ } For each $f$ we have  $E_\omega(f)\in L^2(\Omega,\d{\mathbb P})$. So there is an $\Omega_f\subseteq\Omega$ with ${\mathbb P}(\Omega_f)=1$ s.t. $E_\omega(f)\in\mathbb C$ for all $\omega\in\Omega_f$. (That is, we can choose a representative of the $L^2$ function which is well defined and finite on a set of full measure.) Given $f_1,\ldots, f_K$ we thus find $\Omega_{f_1},\ldots,\Omega_{f_K}$, all of full measure, so that $\sum_{j=1}^k E_\omega(f_j)$ is well defined and finite for all $\omega\in\cap_{j=1}^K\Omega_{f_j}$, again a set of full measure. The latter sum then defines again an element in $L^2(\Omega,\d{\mathbb P})$. In this sense, we can form finite (or countably infinite) linear combinations of $E_\omega(f_j)$. The set of $\omega$ of full measure on which (E3) above holds generally depends on the functions $f_j$.

\bigskip

\noindent
{\bf Construction of $\chi_\omega(f)$.\ }
Let $f\in L^2({\mathbb R}^d, \d k)$ be complex valued. The It\^o stochastic integral 
$$
\int_{{\mathbb R}^d} \d B^\omega(k) f(k)
$$
is a random variable on a probability space $L^2(\Omega,\d{\mathbb P})$. It is constructed for complex random variables as for real ones, see e.g.  \cite{Oksendal}. Let 
\begin{equation}
\label{simplef}
\phi(k) = \sum_{\mu=1}^M z_\mu \phi_{I^\mu_1}(k_1)\cdots \phi_{I^\mu_d}(k_d)
\end{equation}
be a finite linear combination, $z_\mu\in\mathbb C$, of indicator functions  $\phi_{I^\mu_1}(k_1)\cdots\phi_{I^\mu_d}(k_d)$, where $k=(k_1,\ldots,k_d)$, $k_j\in\mathbb R$ and $\phi_{I^\mu_j}$ is the indicator function of the interval $I^\mu_j=[a^\mu_j,b^\mu_j)\subset \mathbb R$. The It\^o integral of $\phi$ is defined to be
$$
\int_{{\mathbb R}^d} \d B^\omega(k) \phi(k) = \sum_{\mu=1}^M z_\mu \prod_{j=1}^d B_j^\omega(b^\mu_j)-B^\omega_j(a^\mu_j),
$$
where $B_j^\omega$, $j=1,\ldots,d$, are $d$ independent Brownian motion random variables on $L^2(\Omega,\d{\mathbb P})$. Using that the Brownian increments $\Delta B^\omega_{j,\mu} := B_j^\omega(b^\mu_j)-B_j^\omega(a^\mu_j)$ are independent for different $j$, as well as for $j$ fixed and different $\mu$, and that they are normal with
$$
{\mathbb E}[\Delta B^\omega_{j,\mu}]=0 \quad \mbox{and}\quad  {\mathbb E}[(\Delta B^\omega_{j,\mu})^2]= b^\mu_j-a^\mu_j,
$$ 
one readily verifies that the following It\^o isometry holds:
\begin{equation}
\label{itoisom}
{\mathbb E}\Big [ \Big| \int_{{\mathbb R}^d}\d B^\omega(k) \phi(k)\Big|^2\Big] =  \int_{{\mathbb R}^d} |\phi(k)|^2\d k.
\end{equation}
This isometry allows us to define the It\^o integral for a general $f\in L^2({\mathbb R}^d,\d k)$ as follows. Take any sequence  $\phi_n(k)$ of simple functions \eqref{simplef} s.t.  $\phi_n\rightarrow f$ in $L^2({\mathbb R}^d, \d k)$, as $n\rightarrow\infty$.  By \eqref{itoisom}, the sequence of random variables $\int_{{\mathbb R}^d}\d B^\omega(k) \phi_n(k)$ is Cauchy in $L^2(\Omega, \d{\mathbb P})$. It thus converges in the $L^2(\Omega, \d{\mathbb P})$-sense to a limit which we take as the definition of  $\int_{{\mathbb R}^d}\d B^\omega(k) f(k)$,
$$
\int_{{\mathbb R}^d}\d B^\omega(k) f(k):= \lim_{n\rightarrow\infty}\int_{{\mathbb R}^d}\d B^\omega(k) \phi_n(k)\qquad\mbox{in $L^2({\mathbb R}^d,\d k)$-sense}.
$$
The It\^o isometry then extends to all $f\in L^2({\mathbb R}^d,\d k)$,
\begin{equation}
\label{itoisom2}
{\mathbb E}\Big [ \Big| \int_{{\mathbb R}^d}\d B^\omega(k) f(k)\Big|^2\Big] =  \int_{{\mathbb R}^d} |f(k)|^2\d k.
\end{equation}

\subsection{Hilbert space representation}
\label{hilbertsection}

In the previous section, we have constructed the following expectation functionals of the Weyl algebra in the infinite volume limit.

\begin{itemize}
\item[$\bullet$]  $E_N(f)$ describing $N$ discrete modes, see Proposition \ref{proposition1}, \fer{6},
\item[$\bullet$]  $\langle E\rangle(f) = \lim_{N\rightarrow\infty}{\mathbb E}[E_{N,\omega}(f)]$ describing the phase averaged functional in the continuous mode limit, where $E_{N,\omega}$ is $E_N$ in which the phases are considered to be i.i.d. random, see Proposition \ref{proposition2}, \fer{001},
\item[$\bullet$]  $E_\omega(f)$ describing independently distributed random phases in the continuous mode limit (with convergence in distribution), see Theorem \ref{thm1} and \eqref{Eomega}. 
\end{itemize}

Since ${\rm Re}\, \chi_\omega(f)$ is normal with mean zero and variance $\sigma^2_\mu(f)$  (see \eqref{newvar}), its characteristic function is given by ${\mathbb E}[\e^{\i t {\rm Re}\, \chi_\omega(f)}] = \e^{-\frac12 t^2 \sigma_\mu(f)^2}$. Therefore, we have from Proposition \ref{proposition2}, \fer{001}, that 
\begin{equation}
\langle E\rangle(f) = {\mathbb E} [ E_\omega(f)] = \int_\Omega \d P(\omega) E_\omega(f).
\label{003}
\end{equation}
Relation \fer{003}  will yield a representation of the state associated to $\langle E\rangle(f)$ as a direct integral over the representations of  $E_\omega(f)$. Furthermore, \eqref{003} shows that taking the expectation $\mathbb E$ and the continuous mode limit $N\rightarrow\infty$ are commuting operations, in the sense that the following diagram commutes,
\begin{equation}
\label{comdiagram}
\begin{array}{ccccc}
 & E_{N,\omega} & \stackrel{{\mathbb E}}{-----\longrightarrow} & {\mathbb E}[E_{N,\omega}] &\\
 & |  & & | &\\
\small N \!\!\!\!\!\! & |  & & | &\!\!\!\!\!\!\!\!\!\!\!\! N\\
 & \downarrow &  &\downarrow & \\
 & E_\omega  & \stackrel{{\mathbb E}}{-----\longrightarrow} & \av{E} & 
\end{array}
\end{equation}
The left down arrow has is a limit in the sense of distributions of random variables.

\medskip

Given a state $\rho$ on a $C^*$-algebra $\frak A$, there exists a unique GNS triple $({\cal H}, \pi,\Psi)$ consisting of a Hilbert space $\cal H$, a representation map $\pi:{\frak A}\rightarrow{\cal B}({\cal H})$  and a normalized vector $\Psi\in\cal H$, such that for all $A\in\frak A$,
\begin{equation}
\label{gns}
\rho(A) = \scalprod{\Psi}{\pi(A)\Psi}.
\end{equation}
Both $E_N$ and $\av{E}$ define states on the Weyl algebra, determined  by $\rho_1(W(f))=E_N(f)$ and $\rho_2(W(f))=\av{E}(f)$, $f\in L^2({\mathbb R}^d,dx)$. 

Consider now the family $E_\omega$.  Being an element of $L^2(\Omega, \d{\mathbb P})$, $E_\omega(f)$  is only well defined (represented by a function with finite complex values)  for $\omega\in\Omega_f\subseteq\Omega$ for some $\Omega_f$ with ${\mathbb P}(\Omega_f)=1$. The range over which $\omega$ varies thus depends on $f$. Therefore, it is not clear that there is {\em any} $\omega\in\Omega$ for which one can define simultaneously $E_\omega(f)$ for all $f\in L^2({\mathbb R}^d,\d x)$. However, we can restrict the range of $f$ to a countable subset of ``test functions" ${\cal D}\subset L^2({\mathbb R}^d,\d x)$. For each $f\in\cal D$, there is an $\Omega_f\subseteq \Omega$, ${\mathbb P}(\Omega_f)=1$, on which $E_\omega(f)$ is well defined, that is, for which one can choose a  representative of the $L^2(\Omega,\d{\mathbb P})$ function which is finite on $\Omega_f$. Being a countable intersection of sets of measure one, the set
$$
\Omega({\cal D}) = \cap_{f\in{\cal D}}\, \Omega_f
$$
has also measure one. Furthermore, for every $\omega\in \Omega({\cal D})$ fixed, $E_\omega(f)$ is well defined for all $f\in\cal D$. Without loss of generality, we may assume that $\cal D$ is a vector space over the countable field ${\mathbb Q}+\i{\mathbb Q}$. (If an original  $\cal D$ is not, then we can consider the set of all finite linear combinations of elements of $\cal D$, with scalars from ${\mathbb Q}+\i{\mathbb Q}$. This is again a countable set and we can take that set for $\cal D$.)

We point out that it follows from the construction of the It\^o integral that $\forall \omega\in \Omega({\cal D})$ and for all $f\in\cal D$, $z\in\mathbb C$, $\chi_\omega(zf)=z\chi_\omega(f)$, understood as an equality of finite complex numbers, for suitable representatives of the $L^2(\Omega,\d{\mathbb P})$ objects. However, full linearity of $\chi_\omega$, using {\em complex} scalars, cannot be guaranteed to hold for all $\omega\in\Omega({\cal D})$. Nevertheless, it follows from the almost everywhere linearity of the It\^o integral (see \cite{Oksendal}) that for every $\omega\in\Omega({\cal D})$, $f\mapsto \chi_\omega(f)$ is $({\mathbb Q}+\i{\mathbb Q})$-linear on $\cal D$. (This means that $\forall \lambda,\mu\in{\mathbb Q}+\i{\mathbb Q}$, $\forall f,g\in{\cal D}$, we have $\chi_\omega(\lambda f+\mu g) = \lambda \chi_\omega(f)+\mu\chi_\omega(g)$ and the last equality holds $\forall\omega\in\Omega({\cal D})$, for suitable (finite) representatives of the $\chi_\omega(\cdot)$.) 

It follows from the above discussion that there exists an $\Omega({\cal D})\subseteq \Omega$ of full measure, such that $\forall\omega\in \Omega({\cal D})$,  $E_\omega$ is an expectation functional on the Weyl algebra with test functions $f\in\cal D$, i.e., the Weyl algebra (over the field $\mathbb C$ of scalars) generated by all $W(f)$, $f\in\cal D$.

\begin{thm}[GNS representation]
\label{repthm} 
\ \ \\

{\rm \bf(1)} The GNS representation associated to $E_N(\cdot)$ is given by
\begin{eqnarray*}
{\cal H} &=& {\cal F}(L^2({\mathbb R}^d,\d x)) \otimes L^2(S^1\times\cdots\times S^1,\d\sigma_1\cdots\d\sigma_N)\\
\pi_N(W(f)) &=& W_{\rm Fock}(f) \otimes \e^{-\i\sum_{j=1}^N \sqrt{2\rho_j} \left\{ \cos\theta_j{\rm Re}\widehat f(k_j) + \sin\theta_j{\rm Im }\widehat f(k_j)\right\}}\\
\Psi&=& \Omega_{\rm Fock}\otimes 1.
\end{eqnarray*}
\qquad \quad \ Here, $\d\sigma$ is the uniform measure on the circle $S^1$ and $1$ is the constant function.

\medskip

{\rm \bf(2)} The GNS representation associated to $\av{E}(\cdot)$ is given by
\begin{eqnarray*}
{\cal H} &=& {\cal H}_\rho\subseteq {\cal F}(L^2({\mathbb R}^d,\d x)) \otimes  {\cal F}(L^2({\mathbb R}^d,\d x))\\
\pi_\rho(W(f)) &=& W_{\rm Fock}(Rf) \otimes W_{\rm Fock}(Tf)\\
\Psi &=& \Omega_{\rm Fock}\otimes \Omega_{\rm Fock}.
\end{eqnarray*}
\qquad \quad \ The maps $R,T: L^2({\mathbb R}^d,\d k)\rightarrow L^2({\mathbb R}^d,\d k)$ are real-linear, given by
\begin{eqnarray}
(Rf)(k) &=& \sqrt{1+\rho(k)}\,  \alpha(k) f(k) +\sqrt{\rho} \, \beta(k) \bar f(k)\\
(Tf)(k) &=& \sqrt{1+\rho(k)}\,  \bar\beta(k) f(k) +\sqrt{\rho} \, \alpha(k) \bar f(k),
\end{eqnarray}
\qquad \quad \ where
\begin{eqnarray}
\alpha  &=& \tfrac12 \left( 1+ |\widehat \mu(2)|\sqrt{\frac{\rho}{1+\rho}}\right)^{1/2} +\tfrac12 \left( 1- |\widehat \mu(2)|\sqrt{\frac{\rho}{1+\rho}}\right)^{1/2}\\
\beta &=& \frac{\overline{\widehat \mu(2)}}{2|\widehat \mu(2)|} 
\left\{ \left( 1+ |\widehat \mu(2)|\sqrt{\frac{\rho}{1+\rho}}\right)^{1/2} -  \left( 1- |\widehat \mu(2)|\sqrt{\frac{\rho}{1+\rho}}\right)^{1/2}\right\}
\end{eqnarray}
{}\qquad \quad \ For $\widehat \mu(2)=0$, we have $\alpha=1$ and $\beta = 0$.

\medskip

{\rm \bf(3)} Let ${\cal D}\subset L^2({\mathbb R}^d, \d x)$ be a test function subspace with associated $\Omega({\cal D})$ satisfying

 \qquad ${\mathbb P} (\Omega({\cal D}))=1$. For every $\omega\in\Omega({\cal D})$,  the GNS representation of  $E_\omega(\cdot)$, as a 

\qquad  functional of the Weyl algebra with test functions in $\cal D$, is given by
\begin{eqnarray*}
{\cal H} &=& {\cal H}_{{\cal D}}\subseteq {\cal F}(L^2({\mathbb R}^d,\d x)) \\
\pi_\omega (W(f)) &=& W_{\rm Fock}(f)\e^{\i{\rm Re}\,\chi_\omega(f)}\\
\Psi &=& \Omega_{\rm Fock}.
\end{eqnarray*}
\end{thm}

A proof of Theorem \ref{repthm} is obtained by direct verification.  In finding the GNS representations, we were of course inspired by the Araki-Woods representations, see \cite{AW} and also \cite{MIdeal} for a textbook explanation. 

\medskip

A representation $\pi$ is called regular if $\alpha\mapsto \pi(W(\alpha f))$ is differentiable at $\alpha=0$, in the strong sense on a dense domain in $\cal H$. For regular representations, one defines the represented Weyl operators
$$
W_\pi(f) = \pi(W(f))
$$
and the represented field operators  by
$$
\Phi_\pi (f) = -i\partial_\alpha|_{\alpha =0} \, \pi(W(\alpha f))
$$
and similarly, the creation and annihilation operators by
\begin{eqnarray}
a^*_\pi(f) &=& 2^{-1/2} \big[ \Phi_\pi(f) - \i\Phi_\pi(\i f)\big],\label{api}\\
a_\pi(f) &=& 2^{-1/2} \big[ \Phi_\pi(f) + \i\Phi_\pi(\i f)\big] = (a^*_\pi(f))^*.\nonumber
\end{eqnarray}
It is not hard to see that the three representations of Theorem \ref{repthm} are all regular, and that the creation and annihilation operators are given as follows.

\begin{prop}[Field and creation operators]
	\label{fieldopsprop}
\ \ \\

{\rm \bf(1)} The field and creation operators associated to $E_N(\cdot)$ are
\begin{eqnarray*}
\Phi_N(f) &=& \Phi_{\rm Fock}(f)\otimes\bbbone_{L^2} -\bbbone_{\cal F}\otimes\sum_{j=1}^N \sqrt{2\rho_j} \left\{ \cos\theta_j{\rm Re}\widehat f(k_j) + \sin\theta_j{\rm Im }\widehat f(k_j)\right\}\\
a^*_N(f) &=& a^*_{\rm Fock}(f)\otimes\bbbone_{L^2} -\bbbone_{\cal F}\otimes \sum_{j=1}^N \sqrt{\rho_j} \e^{-\i\theta_j} \widehat f(k_j).
\end{eqnarray*}

{\rm \bf(2)} The field operators associated to $\av{E}(\cdot)$ are
\begin{eqnarray*}
\Phi_\rho (f) &=& \Phi_{\rm Fock}(Rf)\otimes\bbbone_{\cal F} +\bbbone_{\cal  F}\otimes W_{\rm Fock}(Tf).
\end{eqnarray*}
\qquad \quad \ Since $R, T$ are only real linear, the creation operators have a somewhat cum-

\qquad \!bersome expression. For $\widehat\mu(2)=0$ it reduces to
\begin{eqnarray*}
a^*_\rho(f) &=& a_{\rm Fock}^*(\sqrt{1+\rho}f)\otimes\bbbone_{{\mathcal F}}  + \bbbone_{{\mathcal F}} \otimes a_{\rm Fock}(\sqrt{\rho} \bar f).
\end{eqnarray*}

{\rm \bf(3)} The field and creation operators associated to $E_\omega$, for all $\omega\in\Omega({\cal D})$ and all 

\qquad \!$f\in\cal D$, are 
\begin{eqnarray*}
\Phi_\omega(f) &=& \Phi_{\rm Fock}(f) + {\rm Re}\, \chi_\omega(f)\\
a_\omega^*(f) &=& a^*_{\rm Fock}(f) +\tfrac{1}{\sqrt{2}}\,  \chi_\omega(f).
\end{eqnarray*}
\end{prop}

\bigskip

Let $f_1,\ldots, f_n\in\cal D$. For every $\omega\in \Omega({\cal D})$, we have (strongly on a dense domain of Hilbert space, for instance on the finite-particle subspace of Fock space \cite{BRI})
\begin{eqnarray*}
\Phi_\omega (f_1)\cdots \Phi_\omega(f_n) &=& (-\i)^n \frac{\partial^n}{\partial \alpha_1\cdots \partial \alpha_n}\big|_{\alpha_1=\ldots = \alpha_n=0} \ W_\omega(\alpha_1f_1)\cdots W_\omega(\alpha_nf_n)\\
&=& (-\i)^n \frac{\partial^n}{\partial \alpha_1\cdots \partial \alpha_n}\big|_{\alpha_1=\ldots = \alpha_n=0} \ \e^{\i P} W_\omega(\alpha_1f_1+\cdots +\alpha_nf_n),
\end{eqnarray*}
where $P$ is a real, deterministic phase depending on the $\alpha_j$ and the $f_j$ (originating from the Weyl canonical commutation relations \eqref{weylccr}).

Denote by $\av{\cdot}_\Psi$ the average in the state defined by the vectors $\Psi$ in the GNS representations (in the appropriate Hilbert space). Relation \eqref{003} then implies that 
\begin{eqnarray}
\lefteqn{
{\mathbb E}[ \av{\Phi_\omega (f_1)\cdots \Phi_\omega(f_n)}_\Psi]}\nonumber \\
&=& (-\i)^n \frac{\partial^n}{\partial \alpha_1\cdots \partial \alpha_n}\big|_{\alpha_1=\ldots = \alpha_n=0} \ \e^{\i P} \av{E}(\alpha_1 f_1+\cdots+\alpha_n f_n) \nonumber \\
&=& (-\i)^n \frac{\partial^n}{\partial \alpha_1\cdots \partial \alpha_n}\big|_{\alpha_1=\ldots = \alpha_n=0} \ \e^{\i P} \av{W_\rho(\alpha_1 f_1+\cdots+\alpha_n f_n)}_\Psi \nonumber \\
 &=& (-\i)^n \frac{\partial^n}{\partial \alpha_1\cdots \partial \alpha_n}\big|_{\alpha_1=\ldots = \alpha_n=0} \  \av{W_\rho(\alpha_1 f_1)\cdots W_\rho(\alpha_n f_n)}_\Psi \nonumber \\
&=& \av{\Phi_\rho(f_1)\cdots \Phi_\rho(f_n)}_\Psi. 
\label{equal}
\end{eqnarray}

Relation \eqref{equal} can also be expressed in terms of creation and annihilation operators as follows. Using \eqref{api} and forming linear combinations of \eqref{equal} we see that 
$$
\E\left[\av{\Phi_\omega(f_1)\cdots \Phi_\omega(f_{n-1}) a^\#_\omega(f_n)}_\Psi\right] = \av{\Phi_\rho(f_1)\cdots \Phi_\rho(f_{n-1}) a^\#_\rho(f_n)}_\Psi,
$$
where $a^\#$ stands for either $a$ or $a^*$. Then we can continue the procedure to see that in the relation \eqref{equal}, the product of field operators can be replaced by a product of arbitrary creation and annihilation operators.

This allows us to show that $n$ point functions can be expressed in terms of two point functions only. For $f_1,\ldots,f_p$ and $g_1,\ldots g_q$ $p+q$ test functions in  $\cal D$, we define the block matrix $Q\in M_{p+q}(\C)$ by
\begin{equation}
\label{defQ}
Q=\begin{pmatrix} A & C^T \cr C & B \end{pmatrix},
\end{equation}
where $A \in M_{p}(\C)$, $B\in M_{q}(\C)$, $C\in M_{q, p}(\C)$ are defined by  
$$
A_{ij}=\widehat\mu(2)\bra \bar f_i|\rho f_j\ket, \ \ B_{ij}=\bar{\widehat\mu}(2)\bra  g_i|\rho \bar g_j\ket, \ \ C_{ij}=\bra g_i|\rho f_j\ket
$$
and where $C^T$ is the transpose of $C$. 

\begin{prop}[Quasifreeness]
\label{prop1}
Let $f_1,\ldots,f_p$ and $g_1,\ldots g_q$ be test functions in  $\cal D$. Then
\begin{eqnarray}
\lefteqn{
{\mathbb E}\big[\bra a^*_{\omega}(f_1)\cdots a^*_{\omega}(f_p)a_{\omega}(g_1)\cdots a_{\omega}(g_q)\ket_\Psi\big]}\nonumber \\
&=& \left\{
\begin{array}{ll}
0 & \mbox{if $p+q$ is odd}\\
\sum_{\pi\in{\frak S}'_n}
Q_{\pi(1) \pi(2)}Q_{\pi(3)\pi(4)} \dots Q_{\pi(n-1)\pi(n)} & \mbox{if $n=p+q$ is even.}
\label{m0}
\end{array}
\right.
\end{eqnarray}
Here, ${\frak S}_n'$ is the set of all permutations of $(1,\ldots,n)$ such that  $\pi(1)<\pi(3)<\cdots<\pi(n-1)$.
In particular, if $\hat \mu(2)=0$, 
\begin{eqnarray*}
{\mathbb E}\big[\bra a^*_{\omega}(f_1)\cdots a^*_{\omega}(f_p)a_{\omega}(g_1)\cdots a_{\omega}(g_q)\ket_\Psi\big]=\left\{
\begin{array}{ll}
0 & \mbox{if $p\neq q$}\\
\sum_{\sigma\in {\mathfrak S}_p}\prod_{j=1}^{p}\bra g_{\sigma(j)}|\rho f_j\ket & \mbox{if $p=q$.}
\end{array}
\right.
\label{m1}
\end{eqnarray*}
\end{prop}
In particular, we have
\bea
\E\big[\bra a^*_{\omega}(f_1)\ket_\Psi\big]  &=& \E\big[\bra a_{\omega}(f_1)\ket_\Psi\big] =0,  \nonumber \\
\E\big[\bra a^*_{\omega}(f_1)a^*_{\omega}(f_2)\ket_\Psi \big]&=&\widehat\mu(2)\scalprod{ \bar f_1}{\rho f_2},  \nonumber \\ \nonumber
\E\big[\bra a_{\omega}(g_1)a_{\omega}(g_2)\ket_\Psi\big]&=&\bar{\widehat\mu}(2)\scalprod{  g_1}{\rho \bar g_2}, \\
\E\big[\bra a^*_{\omega}(f_1)a_{\omega}(g_1)\ket_\Psi\big]&=&\scalprod{g_1}{\rho f_1},\nonumber\\
\E\big[\bra a_{\omega}(g_1) a^*_{\omega}(f_1)\ket_\Psi\big]&=&\scalprod{g_1}{(\rho +1)f_1}.
\nonumber
\eea
Note that the functional \eqref{m0} is not gauge invariant unless $\widehat \mu(2)=0$. I.e., it has nonzero average of products with unequal numbers of creation and annihilation operators if $\widehat\mu(2)\neq 0$.

\subsection{Dynamics}

\subsubsection{Reservoir dynamics}

The dynamics on the Weyl algebra is given by a Bogoliubov transformation on the functions $f\in L^2({\mathbb R}^d,\d k)$,
$$
f\mapsto \e^{\i t\varepsilon}f,
$$
where $\varepsilon=\varepsilon(k)$ is a real function of $k\in{\mathbb R}^d$. As an example, for photons, $\varepsilon(k)=|k|$. The dynamics of the three infinite-volume expectation functionals is given as follows. For $N$ discrete modes, \eqref{6}, 
\begin{equation}
E_N(\e^{\i t\varepsilon} f) = E_{{\rm Fock}}(f)\  \e^{ \i\, {\rm Re}\sum_{j=1}^N  \e^{-\i\{ \theta_{j}- t \varepsilon(k_j)\}} \sqrt{2\rho_j}\ \widehat f(k_j)},
\label{6'}
\end{equation}
where we note that $E_{\rm Fock}(\e^{\i  t \varepsilon}f) =E_{\rm Fock}(f)$ for all $t$ and all $f$. The dynamics has thus the effect of shifting the phases associated to the coherent states. 

The phase-averaged expectation functional \eqref{001} evolves according to 
\begin{equation}
\label{001'}
\av{E}(\e^{\i t\varepsilon} f) = E_{\rm Fock}(f) \, \e^{-\frac12\sigma_\mu(\e^{\i t\varepsilon} f)^2},
\end{equation}
where
\begin{equation}
\label{newvar'}
\sigma_\mu (\e^{\i t\varepsilon} f)^2= \int_{{\mathbb R}^d} \rho(k)\Big( |f(k)|^2 +{\rm Re} \, \{\e^{2\i t\varepsilon} \widehat \mu(2) \widehat f(k)^2\} \Big)\d k.
\end{equation}
It follows from the Riemann-Lebesgue Lemma that 
\begin{equation}
\label{RLB}
\lim_{t\rightarrow\infty} \sigma_\mu(\e^{\i t\varepsilon}f)^2 =\|\sqrt{\rho}\widehat f\|^2_2.
\end{equation}
To see how to derive \eqref{RLB} in the situation where $\varepsilon=\varepsilon(|k|)$ is an invertible function of $|k|$, with inverse $|k|=v(\epsilon)$, we use spherical coordinates in ${\mathbb R}^d$,
\begin{equation}
\label{deriv}
\int_{{\mathbb R}^d} \e^{2\i t\epsilon} \rho(k)\widehat f(k)^2 
=\int_0^\infty  \e^{2\i t y}F(y)\d y,
\end{equation}
where $F(y) = v(y)^{d-1}v'(y) \int_{S^{d-1}} J(\vec{\theta}) \  \rho(v(y),\vec{\theta}) \widehat f(v(y),\vec{\theta})^2\d\vec{\theta} $. Then the ordinary Riemann Lebesgue Lemma asserts \eqref{RLB} provided $F\in L^2({\mathbb R}_+,\d y)$.

Therefore, the infinite volume state converges, for large times, to the state determined by the uniform phase distribution ($\d\mu(\theta) = \frac{\d\theta}{2\pi}$). Note that any phase distribution $\mu$ satisfying $\widehat\mu(2)=0$ determines also that same state.

The random phase expectation functional $E_\omega$, \eqref{mf8} satisfies
\begin{equation}
E_\omega(\e^{\i t\varepsilon} f)= E_{\rm Fock}(f) \,\e^{\i \,{\cal N}_\omega(0,\sigma_\mu(\e^{\i t\varepsilon} f)^2)}.
\label{mf8'}
\end{equation}
Due to \eqref{RLB}, 
$$
{\cal N}_\omega(0,\sigma_\mu(\e^{\i t\varepsilon}f)^2)\ \stackrel{\mathcal D}{\longrightarrow}\  {\cal N}_\omega(0,\|\sqrt{\mu}\widehat f\|^2_2), \quad t\rightarrow\infty.
$$
Therefore, the random infinite-volume state converges to the the random infinite volume state with uniform phase distribution in the limit of large times. This convergence in the sense of distributions of random variables.

We have shown the following result.

\begin{prop}[Phase uniformization under reservoir dynamics]
\label{propconv}
Let $\mu$ be a phase distribution satisfying $\widehat\mu(1)=0$. Given any $f\in L^2({\mathbb R}^d,\d x)$, we have, as $t\rightarrow\infty$,
\begin{eqnarray}
\av{E}(\e^{\i t\varepsilon}f) \ {\longrightarrow}\   \av{E}_{\rm unif}(f)\label{con1}\\
E_\omega(\e^{\i t\varepsilon}f) \ \stackrel{\mathcal D}{\longrightarrow}\  E_{\omega,{\rm unif}}(f) \label{con2}.
\end{eqnarray}
The convergence in \eqref{con2} is in distribution of random variables. Here, $\av{E}_{\rm unif}(\cdot)$ and $E_{\omega,{\rm unif}}(\cdot)$ are the expectation functionals \eqref{001} and \eqref{Eomega} in which the phase distribution is uniform, $\d\mu(\theta)=\frac{\d\theta}{2\pi}$.
\end{prop}

{\em Remark.\ } All measures $\d\mu(\theta)$ satisfying $\widehat \mu(1)=0$ and having the same value of $\widehat\mu(2)$ give the same expectation functional $\av{E}$ and $E_\omega$ (see Proposition \ref{proposition2} and Theorem \ref{thm1}).

\subsubsection{Coupling to an open quantum system}
\label{opensystdyn}

We consider an $N$-dimensional quantum system in contact with the reservoir of coherent states in which the phases are uniformly randomly distributed. The Hilbert space of pure states of the system is ${\mathbb C}^N$, that of the reservoir is the GNS space given in point  (3) of Theorem \ref{repthm}. The system dynamics is generated by a self-adjoint Hamiltonian with energy levels $e_1,\ldots,e_N$,
$$
H_\s={\rm diag}(e_1,\ldots,e_N).
$$
The state of the reservoir is invariant under its own dynamics by Proposition \ref{propconv}.  The dynamics is implemented as
$$
\pi_\omega(W(\e^{\i t\varepsilon} f)) =W_{\rm Fock} (\e^{\i t\varepsilon} f) \e^{\i{\rm Re}\chi_\omega(f)} = \e^{\i tH_\r} \pi_\omega(W(f))\e^{-\i tH_\r},
$$
where the reservoir Hamiltonian is 
$$
H_\r =\d\Gamma(\varepsilon).
$$
The uncoupled dynamics is therefore given by the Hamiltonian
$$
H_0 = H_\s\otimes\bbbone_\r + \bbbone_\s\otimes H_\r.
$$
To define a coupled dynamics between the system and the reservoir, one proceeds as follows. The free dynamics is given by the group of $*$automorphisms $\alpha_0^t$ on the algebra of observables ${\frak A}={\cal B}({\mathbb C}^N)\otimes{\cal W}$ (where $\cal W$ is the Weyl algebra), defined by 
$$
\alpha_0^t(A_\s\otimes W(f)) = \e^{\i t H_\s}A_\s\e^{-\i t H_\s}\otimes W(\e^{\i t\varepsilon}f).
$$
Then one defines a coupled dynamics by specifying an interaction operator $V\in\frak A$ and using the  Dyson series
\begin{equation}
\label{dyson}
\alpha^t(A) = \alpha_0^t(A) +\sum_{n\geq 1}\int_0^t\d t_1\cdots\int_0^{t_n-1} \d t_n \ [\alpha_0^{t_n}(V), [ \cdots [ \alpha_0^{t_1}(V),\alpha_0^t(A)] \cdots ] ].
\end{equation}
The series converges in the topology of $\frak A$ and defines the interacting dynamics $\alpha^t$, again a grop of $*$automorphisms on $\frak A$. Applying the representation map $\pi_\omega$ (more precisely, $\bbbone_\s\otimes\pi_\omega$) to \eqref{dyson}, we obtain
\begin{eqnarray}
\lefteqn{
\pi_\omega(\alpha^t(A)) = \tau_0^t(\pi_\omega(A))}\label{tau}\\
&& +\sum_{n\ge 1}\int_0^t\d t_1\cdots\int_0^{t_n-1} \d t_n \ [\tau_0^{t_n}(\pi_\omega (V)), [ \cdots [ \tau_0^{t_1}(\pi_\omega(V)),\tau_0^t(\pi_\omega(A))] \cdots ] ],
\nonumber
\end{eqnarray}
where
$$
\tau_0^t (\cdot ) = \e^{\i tH_0} (\cdot )\e^{-\i t H_0}.
$$
The right side of \eqref{tau} defines a group of $*$automorphisms on the represented algebra of observables which is generated by the self-adjoint operator
$$
H = H_0 + \pi_\omega(V),
$$
acting on ${\mathbb C}^N\otimes{\cal F}(L^2({\mathbb R}^d,\d x))$. From physical considerations, one would like to take $V=G\otimes \Phi(g)$, where $G$ is selfadjoint and $\Phi(g)$ is a field operator. Of course, this $V$ does not belong to $\frak A$ and the above construction cannot be carried out. Nevertheless, one can ``regularize" the interaction by introducing $V_\eta$, depending on a small parameter $\eta$, such that $V_\eta\in\frak A$ and in any regular representation $\pi$ of the algebra $\frak A$,  $\pi(V_\eta)\rightarrow G\otimes\Phi_\pi$, as $\eta\rightarrow 0$ (strongly on a dense domain). One can then, for $\eta>0$, carry out the above construction and finally remove $\eta$ once placed in a representation. Such a procedure is decribed in \cite{FM} --  and other approaches are possible. The dynamics of the coupled system is thus generated by the Hamiltonian
\begin{equation}
\label{hamilt}
H = H_0 + G\otimes\Phi_\omega(g) = H_0 +G\otimes\big( \Phi_{\rm Fock}(g) +{\rm Re}\chi_\omega(g)\big),
\end{equation}
acting on ${\mathbb C}^N\otimes{\cal F}(L^2({\mathbb R}^d,\d x))$. 

We consider an {\em energy conserving} (non-demolition) interaction \cite{JZKGKS, PSE} between the system and the reservoir, which consists in taking an operator $G$ that commutes with $H_\s$,
$$
G={\rm diag} (g_1,\ldots,g_N).
$$
Such models are used to investigate ``phase decoherence" of the small system.

The initial system-reservoir state is disentangled,  given by a density matrix 
$$
P_0 = \rho_\s\otimes|\Omega\rangle\langle\Omega|,
$$
acting on the Hilbert space ${\mathbb C}^N\otimes {\cal F}(L^2({\mathbb R}^d,\d x))$. Here, $\rho_\s$ is an arbitrary intial system  density matrix and the reservoir is in the state $\Omega$, which represents the infinitely extended continuous mode coherent state with uniformly distributed phases. The state of the coupled system at time $t$ is given by 
$$
P(t) = \e^{-\i t H} P_0 \,\e^{\i tH}.
$$
Taking the partial trace over the reservoir Hilbert space yields the reduced system density matrix,
$$
\rho_\s(t) = {\rm Tr}_\r  P(t).
$$
We denote its matrix elements in the energy eigenbasis $\{\varphi_j\}_{j=1}^N$  (with $H_\s\varphi_j = e_j\varphi_j$) by 
\begin{equation}
\label{matel}
\rho_{k,l}(t)= \scalprod{\varphi_k}{\rho_\s(t)\varphi_l} = {\rm Tr} \,P(t) |\varphi_l\rangle\langle\varphi_k|.
\end{equation}
As $[H_\s,G]=0$ the populations (diagonal matrix elements) are time-independent. The off-diagonal ones exhibit time decay (``phase decoherence"). For the energy conserving model at hand, the matrix elements \eqref{matel} can be evaluated exactly. The calculation yields (see Appendix D of \cite{MSB})
\begin{eqnarray}
\rho_{k,l}(t) &=& \e^{-\i t(e_k-e_l)} \e^{-\i t(g_k-g_l) {\rm Re}\chi_\omega(g)}\label{deco1}\\
&& \times \e^{\frac\i2(g^2_k-g^2_l) \scalprod{g}{\frac{\sin(\varepsilon t)-\varepsilon t}{\varepsilon}g}}
\e^{-\frac12 (g_k-g_l)^2 \scalprod{g}{\frac{1-\cos(\varepsilon t)}{\varepsilon^2}g}}\label{deco2}\\
&& \times\rho_{k,l}(0).\nonumber
\end{eqnarray}
The contribution on the right side of \eqref{deco1} is given by the free dynamics and by a random ``renormalization" of the system energy due to the interaction with the coherent bath (coming from the term $G\otimes {\rm Re}\chi_\omega(g)\bbbone_\r$ in the Hamiltonian \eqref{hamilt}). The two factors \eqref{deco2} are the same as if the system was coupled to a free bose gas in equilibrium at zero temperature. Therefore, coherent states character of the reservoir is encoded entirely in the part $\e^{-\i t(g_k-g_l) {\rm Re}\chi_\omega(g)}$. The expectation of this oscillating factor is the characteristic function of the random variable ${\rm Re}\chi_\omega(g)\sim{\cal N}_\omega(0,\|\sqrt{\rho}g\|^2_2)$,
$$
\E\left[ \e^{-\i t(g_k-g_l) {\rm Re}\chi_\omega(g)}\right] = \e^{-\frac{t^2}{2} (g_k-g_l)^2\|\sqrt{\rho}  g\|^2_2 }.
$$
This shows that the averaged (reduced system) density matrix $\E[\rho_\s(t)]$ acquires {\em Gaussian time-decay} of off-diagonals at all times, due to the coupling with the coherent reservoir, namely
\begin{equation}
\big| \E[\rho_{k,l}(t)] \big| = \e^{-\frac{t^2}{2} (g_k-g_l)^2\|\sqrt{\rho}  g\|^2_2 } \, \e^{-\frac12 (g_k-g_l)^2\Gamma(t)}\, |\rho_{k,l}(0)|,
\label{deco3}
\end{equation}
with
\begin{equation}
\label{gammat}
\Gamma(t) =  \scalprod{g}{\frac{1-\cos(\varepsilon t)}{\varepsilon^2}g} = 2\scalprod{g}{\frac{\sin^2(\varepsilon t/2)}{\varepsilon^2}g}.
\end{equation}
For small times, $\Gamma(t) \sim \tfrac12 t^2\|g\|^2_2$ is quadratic in time, but for large $t$, its behaviour as a function of $t$ depends on the infrared behaviour of the form factor $g$. For instance, in $d=3$ dimensions and for $\varepsilon(k)=|k|$,
$$
\Gamma(t) =2\int_0^\infty |k|^2\d|k| \int_{S^2}\d\Sigma |g(|k|,\Sigma)|^2 \frac{\sin^2(|k| t/2)}{|k|^2} \sim \frac{\pi t}{2}  \lim_{r\rightarrow 0_+} r^2 \int_{S^2}\d\Sigma |g(r,\Sigma)|^2,
$$
assuming that the latter limit exists and is non-vanishing, meaning that $|g(r,\Sigma)|\sim r^{-1}$ for small $r$. Note also that for $p>-1/2$, we have $\lim_{t\rightarrow\infty}\scalprod{g}{\cos(\varepsilon t)/\varepsilon^2g}=0$ by the Riemann-Lebesgue lemma, so that $\lim_{t\rightarrow\infty} \Gamma(t)=\|g/\varepsilon\|^2_2$. For this infra-red behaviour of the form factor, the coupling to the (zero temperature) reservoir does not induce (complete) decoherence, but the coupling to the coherent reservoir does.

\section{Proofs}

\subsection{Proof of Proposition \ref{proposition1}}

We calculate the limit $L\rightarrow\infty$ of
\begin{equation}
E_N^\Lambda(f)= \scalprod{\widehat\Psi}{W(\widehat{f})\widehat\Psi},
\label{3}
\end{equation}
where $W(\widehat f) = \e^{\i \Phi(\widehat f)}= \e^{\frac{\i}{\sqrt{2}} \sum_k  \widehat{f}_k a^*_k +\overline{\widehat{f}}_k a_k}$ is the Weyl operator in momentum space.  The coherent state \fer{cohstate} has the form
$$
\widehat\Psi = W(\widehat g)\widehat\Omega,\quad\mbox{with}\quad \widehat g_k = \left\{
\begin{array}{ll}
0 & \mbox{for $k\neq k_j'(L)$}\\
-\i\sqrt{2}\alpha_j(L) & \mbox{for $k=k_j'(L)$}
\end{array}
\right.
$$
and where the momenta satisfy $\lim_{L\rightarrow\infty}k'_j(L) = k_j$ and $\alpha_j(L)$ is given in \eqref{alpha}. Using the canonical commutation relations $W(\widehat f)W(\widehat g)=\e^{-\frac{\i}{2}{\rm Im}\scalprod{\widehat f}{\widehat g}} W(\widehat f+\widehat g)$ yields
$$
E_N^\Lambda(f) = \scalprod{\widehat\Omega}{W(-\widehat g) W(\widehat f) W(\widehat g) \widehat \Omega}
=E_{\rm Fock}(f)\, \e^{\i\sqrt{2}{\rm Re}\,\sum_{j=1}^N \bar\alpha_j(L) \widehat f_{k'_j(L)}}.
$$
Combining \eqref{alpha} with the formula \eqref{1} gives
$$
\sum_{j=1}^N \bar\alpha_j(L)\widehat f_{k'_j(L)} = \sum_{j=1}^N\sqrt{\rho_j}\e^{-\i\theta_j}\int_\Lambda \e^{-\i k'_j(L) x} f(x)\d x,
$$
which converges to $\sum_{j=1}^N\sqrt{\rho_j}\e^{-\i\theta_j} \int_{{\mathbb R}^d}\e^{-\i k_j x}f(x)\d x$ in the limit $L\rightarrow\infty$. \hfill \qed

\subsection{Proof of Proposition \ref{cltprop}}
\label{sectbilling}

The mechanism of the proof of Proposition \ref{cltprop} is the following. Since the $\xi_j$, defined in \eqref{xiomega}, are independent with mean zero and variance 
$$
{\mathbb E}[\xi_j^2] = (2R)^d \rho(k_j) \Big( |\widehat f(k_j)|^2 +{\rm Re} \{ \widehat\mu(2) \widehat f(k_j)^2\} \Big),
$$
where $\widehat \mu(2)$ is given in \eqref{newvar}, a version of the central limit theorem says that 
\begin{equation}
\frac{N^{-d/2} \sum_{j\in\{1,\ldots,N\}^d} \xi_j(\omega)}{\sqrt{N^{-d}\sum_{j\in\{1,\ldots, N\}^d} {\mathbb E}[\xi_j^2]}\ \ } \ \stackrel{\mathcal D}{\longrightarrow}\ {\cal N}_\omega(0,1),\quad \mbox{as $N\rightarrow\infty$}.
\label{mf5}
\end{equation}
Here, ${\cal N}_\omega(0,1)$ is the standard normal with mean zero and variance one. We point out that the $\xi_j$ are not identically distributed (only the $\theta_j$ are). We give a proof of \eqref{mf5} here below. The denominator in \fer{mf5} has the limit $\sigma_\mu(f)$ given in \eqref{newvar}, as $N\rightarrow\infty$. 
Hence \fer{mf5} implies
$$
N^{-d/2} \sum_{j\in\{1,\ldots,N\}^d} \xi_j(\omega) \ \stackrel{\mathcal D}{\longrightarrow}\ \sigma_\mu(f) \,{\cal N}_\omega(0,1) = {\cal N}_\omega\big(0,\sigma_\mu(f)^2\big),\quad \mbox{as $N\rightarrow\infty$}.
$$

\medskip

We now prove \eqref{mf5}. Define
\be
s_N^2:=\sum_{j\in\{1,\ldots,N\}^d}  \E[\xi_j^2(\omega)] = 
(2R)^d \sum_{j\in\{1,\ldots,N\}^d} \rho(k_j) \Big( |\widehat f(k_j)|^2 +{\rm Re} \{ \widehat\mu(2) \widehat f(k_j)^2\} \Big).
\ee
The Central Limit Theorem (see Theorem 27.3 in \cite{Billingsley}) says that 
$$
\frac{1}{s_{N}}\sum_{j\in\{1,\ldots,N\}^d}  \xi_j(\omega)\ \stackrel{\mathcal D}{\longrightarrow}\ {\mathcal N}_\omega(0,1), \ \ \mbox{as} \ N\ra \infty,
$$
provided that the following Lyapounov condition holds, 
\be\label{condclt}
\lim_{N\ra \infty}\sum_{j\in\{1,\ldots,N\}^d} \frac{\E[\, |\xi_j(\omega)|^{2+\delta}]}{s_N^{2+\delta}}=0 
\ee
for some $\delta>0$.  We have
$$
\sum_{j\in\{1,\ldots,N\}^d } \E[\, |\xi_j(\omega)|^{2+\delta}] \le 
(2R)^{d(2+\delta)/2}\sum_{j\in\{1,\ldots,N\}^d } (2\rho(k_j))^{(2+\delta)/2} |\widehat f(k_j)|^{2+\delta} \sim N^d
$$
as well as
$$
s_N^{2+\delta}\le (2R)^{d(2+\delta)/2}\Big\{ \sum_{j\in\{1,\ldots, N\}^d} 2\rho(k_j)\, |\widehat f (k_j)|^2
\Big\}^{(2+\delta)/2} \sim  N^{d(2+\delta)/2}.
$$
Therefore \eqref{condclt} holds since the sum is of the order $N^{-d\delta/2}$. \hfill \qed

\subsection{Proof of Proposition \ref{prop1}}

Consider the complex valued random variable 
\bea\label{rancorr}
&&\bra a^*_{\omega}(f_1)\cdots a^*_{\omega}(f_p)a_{\omega}(g_1)\cdots a_{\omega}(g_q)\ket_\Psi=\nonumber\\
&& \hspace{3cm} \frac1{\sqrt2^{p+q}}\chi_\omega(f_1)\cdots \chi_\omega(f_p)\overline{\chi_\omega(g_1)}\cdots \overline{\chi_\omega(g_q)}.
\eea
{}For functions $f_1,\cdots, f_p$, $g_1, \cdots, g_q$ in $L^2$ and complex numbers $z_1,\cdots, z_p$, $w_1, \cdots, w_q$, define
\begin{equation}
\label{phidef}
\Phi(z,w)=\E\Big[\exp\big\{\sum_jz_j\chi_\omega(f_j)+\sum_k w_k\overline{\chi_{\omega}(g_k)}\big\} \Big].
\end{equation}
The derivative at the origin yields the sought for expectation value,
\be\label{exprod}
\partial^{p+q}_{z_1 \cdots z_p w_1 \cdots w_q}|_{(0,0)}\Phi(z,w)=\E\Big[\chi_\omega(f_1)\cdots \chi_\omega(f_p)\overline{\chi_\omega(g_1)}\cdots \overline{\chi_\omega(g_q)}\Big].
\ee
We have
\bea\label{scsc}
&&\sum_jz_j\chi_\omega(f_j)+\sum_k w_k\overline{\chi_{\omega}(g_k)}=
\chi_\omega(\sum_jz_jf_j)+\overline{\chi_{\omega}(\sum_k \overline{w_k}g_k)}\\  \nonumber
&&\hspace{3cm}=\chi_{\omega,1}\Big(\sum_jz_jf_j+\sum_k \overline{w_k}g_k\Big)+\i \chi_{\omega,2}\Big(\sum_jz_jf_j-\sum_k \overline{w_k}g_k\Big),
\eea
where $\chi_{\omega,1}(\cdot)$ and $\chi_{\omega,2}(\cdot)$ are the real and imaginary parts of $\chi_\omega(\cdot)$. 

Consider the characteristic function of $(\chi_{\omega,1}(f),\chi_{\omega,2}(g))$. For any real $\tau_1, \tau_2$, the random variable $\tau_1\chi_{\omega,1}(f)+\tau_2 \chi_{\omega,2}(g)$ is 
normal with zero expectation value and 
\bea\label{gencar}
\lefteqn{
\E\big[\e^{\i (\tau_1\chi_{\omega,1}(f)+\tau_2 \chi_{\omega,2}(g))} \big] }\\
\nonumber &=&\exp\left\{-\frac12\left[\tau_1^2\mbox{var}\chi_{\omega,1}(f)+\tau_2^2 \mbox{var}\chi_{\omega,2}(g)+2\tau_1\tau_2\mbox{cov}(\chi_{\omega,1}(f), \chi_{\omega,2}(g))\right]\right\}.
\eea
Note that $\tau_1, \tau_2$ can be taken here as arbitrary complex numbers, since the distribution of $\chi_{\omega,j}(\cdot)$, $j=1,2$, is Gaussian.

\begin{lem}\label{struphi}
With the notations above,
\bea\label{pzw}
\Phi(z,w)&=&\exp\left\{ 2 \sum_{j,  k}  z_jw_k\bra g_k | \rho f_j\ket \right. \nonumber \\
&+&\left.  \hat\mu(2)\sum_{j,j'}z_jz_{j'}\bra \bar f_{j'}| \rho f_j\ket+ \bar{\hat\mu}(2)\sum_{k,k'}w_kw_{k'}\bra g_{k'}|\rho \bar g_k\ket\right\}.
\eea
\end{lem}

{\bf Proof:}
According to \eqref{phidef} and \eqref{scsc}, $\Phi$ is given by \eqref{gencar} with $\tau_1=-\i$, $\tau_2 = 1$ and $f=\sum_j z_j f_j +\sum_k \bar w_k g_k$, $g=\sum_jz_jf_j -\sum_k\bar w_kg_k$. Set $a= \sum_{j=1}^pz_jf_j$ and $b=\sum_{k=1}^q \overline{w_k}g_k$. Using (\ref{mf10}), (\ref{s1}) and (\ref{s2}), we compute
\be\nonumber
\mbox{var}\chi_{\omega,1}(a+b)-\mbox{var}\chi_{\omega,2}(a-b)=\int \rho(k) \left\{2(a\bar b +\bar a b)+\hat\mu(2)(a^2+b^2)+\overline{\hat\mu(2)(a^2+b^2)}\right\}dk,
\ee
and
\be\nonumber
\mbox{cov}(\chi_{\omega,1}(a+b),\chi_{\omega,2}(a-b))=\frac{1}{2i}\int \rho(k)\left\{2(a\bar b -\bar a b)+\hat\mu(2)(a^2-b^2)-\overline{\hat\mu(2)(a^2-b^2)}\right\}dk,
\ee
from which (\ref{pzw}) follows. \hfill\qed

\smallskip

We use the notation $t=(z,w)\in \C^{p}\times \C^q=\C^n$ and write accordingly $\Phi(t)\equiv \Phi(z,w)$. Lemma \ref{struphi} shows that $\Phi(t)=e^{q(t)}$, where $q(t)=\sum_{j,k \in \{1,\dots, n\}} t_j t_k Q_{jk}$ is a quadratic form, with corresponding matrix $Q\in M_{p+q}(\C)$
defined by (\ref{defQ}).
Equation \eqref{exprod}  takes the form
\begin{equation}
\label{sp1}
\E\Big[\chi_\omega(f_1)\cdots \chi_\omega(f_p)\overline{\chi_\omega(g_1)}\cdots \overline{\chi_\omega(g_q)}\Big] = \partial^n_{t_1,\ldots,t_n}|_{0} \ \sum_{\ell\ge 0} \frac{1}{\ell !} \big( \sum_{j,k=1}^n t_j t_k Q_{jk}\big)^\ell.
\end{equation}
The right hand side is not zero only if $n$ is even and only the term $\ell =n/2$ in the series does not vanish, so that we have
\begin{equation}
\label{ps3}
\E\Big[\chi_\omega(f_1)\cdots \chi_\omega(f_p)\overline{\chi_\omega(g_1)}\cdots \overline{\chi_\omega(g_q)}\Big] = \partial^n_{t_1,\ldots,t_n}|_{0} \ \frac{1}{(n/2)!}\big( \sum_{j,k=1}^n t_j t_k Q_{jk}\big)^{n/2}.
\end{equation}
 The power $n/2$ of the double sum over $j,k$ in the right hand side of \eqref{ps3}  is an $n$-fold sum over values  $t_s$, $s=1,\ldots,n$ and each summand has a factor of the form $t_{j_1}t_{k_1}\cdots t_{j_{n/2}}t_{k_{n/2}}$. The only terms in that multiple sum that have a nonzero derivative are those in which $(j_1,k_1,\ldots,j_{n/2},k_{n/2})$ are a permutation of $(1,2,\ldots,n)$. Therefore,
\begin{equation}
\label{ps4}
\E\Big[\chi_\omega(f_1)\cdots \chi_\omega(f_p)\overline{\chi_\omega(g_1)}\cdots \overline{\chi_\omega(g_q)}\Big] = \frac{1}{(n/2)!} \sum_{\pi\in{\frak S}_n} Q_{\pi(1)\pi(2)}\cdots Q_{\pi(n-1)\pi(n)}.
\end{equation}
Each term in the sum of \eqref{ps4} has the same value upon permuting the $n/2$ different factors $Q$, so we arrive at \eqref{m0}. This concludes the proof of Proposition \ref{prop1}.\hfill \qed

\subsection{Simultaneous continuous mode and infinite volume limits}   

\label{sameresultsection}

We consider the one-dimensional case, $d=1$. The discussion for general $d$ is analogous. Suppose we want an infinite volume coherent state with modes distributed in an interval $[a,b]\subset\mathbb R$. 
The finite volume state is
$$
\e^{\sum_{k\in[a,b]} \alpha_{k} a^*_{k} - \overline{\alpha}_{k} a_{k}}\widehat \Omega.
$$
The discrete modes in $[a,b]$ are $k_j=a+j\frac{\pi}{L}(b-a)$ and the sum in the exponent is 
$$
\sum_{j=0}^{L/\pi} \alpha_{k_j} a^*_{k_j} - \overline{\alpha}_{k_j} a_{k_j}.
$$
This gives the expectation functional (see \fer{6})
\begin{equation}
E_L( g) = E_{{\rm Fock}}(g)  \e^{\i \sqrt{2} {\rm Re}\sum_{j=1}^{L/\pi}\overline{\alpha}_{k_j}\, \widehat g_{k_j}} =  E_{{\rm Fock}}(g)  \e^{\i \sqrt{2} {\rm Re}\frac{1}{\sqrt{L}}\sum_{j=1}^{L/\pi}\overline{\alpha}_{k_j}\, \widehat g(k_j)}. 
\label{8}
\end{equation}
The total particle density is $\rho = \frac{1}{2L}\sum_{j=0}^{L/\pi} |\alpha_{k_j}|^2 = \sum_{j=0}^{L/\pi}  \rho_{k_j}$. Let $\rho(k)$ be the density distribution, i.e., $\int_I\rho(k)\d k$ is the density of particles having moment in the interval $I\subset [a,b]\subset\mathbb R$. Then $\rho(k_j)=\rho_{k_j}/\Delta k_j=\rho_{k_j}L/\pi$ and hence $\rho = \sum_{j=0}^{L/\pi}\rho(k_j)\Delta k_j$, so $\rho=\int_a^b\rho(k)\d k$ in the limit $L\rightarrow\infty$. The link between $\alpha_{k_j}$ and $\rho(k_j)$ is
$$
\alpha_{k_j} = \sqrt{2L\rho_{k_j}}\ \e^{\i\theta_{k_j}} = \sqrt{2L\rho(k_j)}\ \e^{\i\theta_{k_j}} \sqrt{\Delta k_j} = \sqrt{2\pi \rho(k_j)}\ \e^{\i\theta_{k_j}},
$$
since  $\Delta k_j = \pi/L$. We use this in \fer{8},
$$
E_L(g) =E_{{\rm Fock}}(g) \  \e^{2\i {\rm Re} \sqrt{\frac{\pi}{L}}\sum_{j=1}^{L/\pi}\sqrt{\rho(k_j)}\ \e^{-\i \theta_{k_j}}\, \widehat g(k_j)}.
$$
The phase behaves like $2\sqrt{L/\pi}\ {\rm Re}\int_a^b \sqrt{\rho(k)}\e^{-\i\theta(k)}\widehat g(k)\d k$  and diverges like $\sqrt{L}$.

\subsection{Rarefied continuous mode limit}

We consider $d=1$. General $d$ are treated in the same way. 
The origin of the non-existence of the continuous mode limit \eqref{mf3}, \eqref{div} is the following. The discrete density $\sqrt{\rho_{k_j}}$ is related to the continuous density distribution $\rho(k_j)$ by $\rho_{k_j} \propto \frac1N\rho(k_j)$, where $1/N$ is the discretization mesh size. The phase of the expectation functional is a sum over the square root of the density, $\sum_{j=1}^N \sqrt{\rho_{k_j}} \widehat g(k_j) = \frac{1}{\sqrt{N}}\sum_{j=1}^N \sqrt{\rho(k_j)} \widehat g(k_j)$, giving a prefactor which is too weak for convergence of the Riemann sum. 

Instead of populating all modes, we may consider a rarefied situation, where, in a given interval $[a,b]$, only a fraction of modes are chosen.  The modes in the entire interval are $k_j=a+j\frac{\pi}{L}(b-a)$, $j=0,\ldots,L/\pi$. Fix an integer $s$ and choose
$$
\alpha_{k_j} = \left\{
\begin{array}{cl}
\alpha_{k_j} &  \mbox{for $j=\ell s$, $\ell=0,\ldots,\frac{L}{s\pi}$}\\
0 & \mbox{for all other $j$}
\end{array}
\right.
$$
The sum in the phase of \fer{8} becomes
$$
\frac{1}{\sqrt{L}} \sum_{j=1}^{L/\pi} \overline{\alpha}_{k_j}\, \widehat g(k_j) = \frac{1}{\sqrt{L}} \sum_{\ell=1}^{L/s\pi} \overline{\alpha}_{a+\ell s\pi(b-a)/L}\, \widehat g(a+\ell s\pi(b-a)/L).
$$
The mesh size in the Riemann sum over $\ell$ is $\Delta k = \frac{s\pi}{L}(b-a)$, and we see that the right side converges, and has a nonzero limit, only if $(\Delta k\sqrt{L})^{-1}$ converges to a nonzero limit. This forces $s\propto\sqrt{L}$. Taking $s=\frac{\sqrt{L}}{\sigma\pi(b-a) }$ for some $\sigma>0$, we obtain
$$
\lim_{L\rightarrow\infty}\frac{1}{\sqrt{L}} \sum_{j=1}^{L/\pi} \overline{\alpha}_{k_j}\, \widehat g(k_j) = \sigma \int_a^b\overline{\alpha}(k)\widehat g(k) \d k.
$$
The infinite volume expectation functional becomes
$$
\lim_{L\rightarrow\infty}E_L(g) = E_{{\rm Fock}}(g) \    \e^{\i \sqrt{2}\sigma {\rm Re}\int_a^b\overline{\alpha}(k)\widehat g(k) \d k }. 
$$
We examine the particle number and density, as $L\rightarrow\infty$. The number of particles is 
$$
\av{N}_L = \sum_{j=0}^{L/\pi} |\alpha_{k_j}|^2 = \sum_{\ell=1}^{L/s\pi} |\alpha_{a+\ell s\pi(b-a)/L}|^2 \sim \sigma \sqrt{L}\int_a^b |\alpha(k)|^2\d k.
$$
(Note, the mesh size in the last Riemann sum is $\Delta k =\frac{s\pi}{L}(b-a)= (\sigma\sqrt{L})^{-1}$.) Therefore, the number of particles grows like $\sigma$ times the square root of the volume. Dividing by the volume, we obtain that the total particle density decays to zero like $\sigma/\sqrt{L}$. We have thus an infinite volume limit state containing infinitely many particles, but having zero particle density  -- a rarefied state.

\bigskip

{\bf Acknowledgements. } The authors are grateful to the Institut Fourier, Grenoble, as well as the Department of Mathematics and Statistics of Memorial University, for supporting this research. M.M. thanks the Natural Sciences and Engineering Research Council of Canada (NSERC) for financial support.

\end{document}